\documentclass[10pt]{article}
\usepackage{amsmath, amsthm}
\usepackage{amscd}
\usepackage{amsfonts}
\usepackage{amssymb}
\usepackage{mathrsfs}

\makeindex

\newtheorem{theorem}{Theorem}[section]
\newtheorem{proposition}[theorem]{Proposition}
\newtheorem{lemma}[theorem]{Lemma}

\theoremstyle{definition}
\newtheorem{definition}[theorem]{Definition}

\newtheorem{remark}[theorem]{Remark}

\def\proof{{\noindent\sc Proof. \quad}}

\def\eproof{{\mbox{}\hfill\qed}\medskip}
\begin{document}

\makeatletter

\renewcommand{\bar}{\overline}

%Other math symbols
\newcommand{\x}{\times}
\newcommand{\<}{\langle}
\renewcommand{\>}{\rangle}
\newcommand{\into}{\hookrightarrow}

%Greek letters
\renewcommand{\a}{\alpha}
\renewcommand{\b}{\beta}
\renewcommand{\d}{\delta}
\newcommand{\D}{\Delta}
\newcommand{\e}{\varepsilon}
\newcommand{\g}{\gamma}
\newcommand{\G}{\Gamma}
\renewcommand{\l}{\lambda}
\renewcommand{\L}{\Lambda}
\newcommand{\n}{\nabla}
\newcommand{\var}{\varphi}
\newcommand{\s}{\sigma}
\newcommand{\Sig}{\Sigma}
\renewcommand{\t}{\theta}
\renewcommand{\O}{\Omega}
\renewcommand{\o}{\omega}
\newcommand{\z}{\zeta}
\newcommand{\balpha}{\boldsymbol \alpha}
\newcommand{\ab}{\alpha_{\bullet}}

%Other macros
\newcommand{\p}{\partial}
\renewcommand{\hat}{\widehat}
\renewcommand{\bar}{\overline}
\renewcommand{\tilde}{\widetilde}

%%%%%%%%
% Some fonts
%%%%%%%

%\newcommand{\fiverm}{\tiny\rm}
\font\eightrm=cmr8
\font\ninerm=cmr9

%%%%%
% The next macros define fonts for reals, rationals, complex,
% integers and natural numbers by \R,\Q,\C,\Z and \N respectively.
% Also, \Ri gives \R to the infinity.
%%%%%

\def\N{\mathbb{N}}
\def\Z{\mathbb{Z}}
\def\R{\mathbb{R}}
\def\Q{\mathbb{Q}}
\def\C{\mathbb{C}}
\def\F{\mathbb{F}}
\def\proj{\mathbb{P}}
\def \Ri{\R^\infty}
\def \Zi{\Z^\infty}
\def \ZRi{\Z^\infty\x\R^\infty}
\def \SZi{\Z^\infty\x S(\R^\infty)}

%%%%%%%
% For writting algorithms
%%%%%%%

\def\algo{\begin{center}
               \begin{minipage}{6in}
               \begin{tabbing}
               \marks}
\def\falgo{\end{tabbing}
                \end{minipage}
                \end{center}}
\def\marks{nn\= nn\= nn\= nn\= nn\= nn\= nn\= \kill}

%%%%%%%
% Some mathematical operators and constants
%%%%%%%

\def\ll{{[\kern-1.6pt [}}
\def\rr{{]\kern-1.4pt ]}}
\def\bll{{\biggl[\kern-3pt \biggl[}}
\def\brr{{\biggr]\kern-3pt \biggr]}}
\def\sg{{\rm sign}\,}
\def\sig{\overline{\rm sign}\,}
\def\absmax{\mathop{\underline{\rm max}}\limits}
\def\rank{{\rm rank}\,}
\def\Im{{\rm Im}}
\def\dist{{\rm dist}}
\def\degree{{\rm degree}\,}
\def\grad{{\rm grad}\,}
\def\size{{\rm size}}
\def\supp{{\rm supp}}
\def\Id{{\rm Id}}
\def\fl{\mathop{\tt fl}}
\def\op{\mathop{\tt op}}
\def\sizem{{\rm size}_\mu}
\def\sizew{{\rm size}_{\rm h}}
\def\cost{{\rm cost}}
\def\card{{\rm card}}
\def\mod{{\rm mod}\;}
\def\trace{{\rm trace}}
\def\Prob{\mathop{\rm Prob}}
\def\length{{\rm length}}
\def\Diag{{\rm Diag}}
\def\diam{{\rm diam}}
\def\arg{{\rm arg}\,}
\def\ot{\leftarrow}
\def\transp{^{\rm T}}
\def\bL{{\bf L}}
\def\bR{{\bf R}}
\def\bfE{{\bf E}}
\def\bd{{\bf d}}
\def\adj{{\rm adj}\:}
\def\cruz{\raise0.7pt\hbox{$\scriptstyle\times$}}
\def\todif{\stackrel{\scriptscriptstyle\not =}{\to}}
\def\nedif{\stackrel{\scriptscriptstyle\not=}{\nearrow}}
\def\sedif{\stackrel{\scriptscriptstyle\not=}{\searrow}}
\def\nodown{\downarrow\mkern-15.3mu{\raise1.3pt\hbox{$\scriptstyle\times$}}}
\def\noto{\to\mkern-20mu\cruz}
\def\notox{\to\mkern-23mu\cruz}
\def\alg{{\hbox{\scriptsize\rm alg}}}
\def\ttl{{\tt l}}
\def\tto{{\tt o}}
\def\ttml{\bar{\tt l}}
\def\ttmo{\bar{\tt o}}
\def\ri{R^\infty}
\def\Oh{{\cal O}}
\def\parts{{\cal P}}
\def\coeff{{\hbox{\rm coeff}}}
\def\Error{{\hbox{\tt Error}}\,}
\def\appleq{\hbox{\lower3.5pt\hbox{$\;\:\stackrel{\textstyle<}{\sim}\;\:$}}}
%\def\appleq{{<\atop{\sim}}}

% Para usar en dibujos con Pictex %%
\def\bolita{\scriptscriptstyle\bullet}

%%%%%%%
% Some abreviations for the bib database
%%%%%%%

\def\JACM{Journal of the ACM}
\def\CACM{Communications of the ACM}
\def\ICALP{International Colloquium on Automata, Languages
            and Programming}
\def\STOC{annual ACM Symp. on the Theory
          of Computing}
\def\FOCS{annual IEEE Symp. on Foundations of Computer Science}
\def\SIAM{SIAM J. Comp.}
\def\SIOPT{SIAM J. Optim.}
\def\BSMF{Bulletin de la Soci\'et\'e Ma\-th\'e\-ma\-tique de France}
\def\CRAS{C. R. Acad. Sci. Paris}
\def\IPL{Information Processing Letters}
\def\TCS{Theoret. Comp. Sci.}
\def\BAMS{Bulletin of the Amer. Math. Soc.}
\def\TAMS{Transactions of the Amer. Math. Soc.}
\def\PAMS{Proceedings of the Amer. Math. Soc.}
\def\JAMS{Journal of the Amer. Math. Soc.}
\def\LNM{Lect. Notes in Math.}
\def\LNCS{Lect. Notes in Comp. Sci.}
\def\JSL{Journal for Symbolic Logic}
\def\JSC{Journal of Symbolic Computation}
\def\JCSS{J. Comput. System Sci.}
\def\JoC{J. Compl.}
\def\MP{Math. Program.}
\sloppy

\bibliographystyle{plain}

%new macros
%%%%%%%%%%%%%%%%%%%

\def\CPRi{{\rm \#P}_{\kern-2pt\R}}

\def\CC{{\mathcal C}}
\def\DD{{\mathcal D}}
\def\NN{{\mathcal N}}
\def\MM{{\mathcal M}}
\def\GG{{\mathcal G}}
\def\UU{{\mathcal U}}
\def\ZZ{{\mathcal Z}}
\def\PP{{\mathscr P}}
\def\AA{{\mathscr A}}
\def\scC{{\mathscr C}}
\def\scE{{\mathscr E}}
\def\sG{{\mathscr G}}
\def\mZ{{\mathcal Z}}
\def\mI{{\mathcal I}}
\def\sH{{\mathscr H}}
\def\sF{{\mathscr F}}
\def\bD{{\mathbf D}}
\def\nrr{{\#_{\R}}}
\def\Sd{{\Sigma_d}}
\def\oPp{\overline{P'_a}}
\def\oP{\overline{P_a}}
\def\oDH{\overline{DH_a^\dagger}}
\def\oH{\overline{H_a}}
\def\ua{\overline{\a}}
\def\Hd{\HH_{\mathbf d}}
\def\Lg{{\rm Lg}}
%%%%%

\def\bx{{\bf x}}
\def\ii{{\'{\i}}}

\def\P{\mathbb P}

\newcommand{\binomial}[2]{\ensuremath{{\left(
\begin{array}{c} #1 \\ #2 \end{array} \right)}}}

\newcommand{\HH}{\ensuremath{\mathcal H}}
\newcommand{\diag}{\mathbf{diag}}
\newcommand{\CH}{\mathsf{CH}}
\newcommand{\Cone}{\mathsf{Cone}}
\newcommand{\SCH}{\mathsf{SCH}}

\newcounter{line}
\newcounter{algorithm}

\newenvironment{algorithm}[3]
{
\addtocounter{algorithm}{1}
{\bf Algorithm \thealgorithm : \sf #1} \\
{\bf Input: }#2 \\
{\bf Output: }#3 \\
\begin{list}{\arabic{line}:}{\usecounter{line}}\setlength{\leftmargin}{3em}
}
{
\end{list}
}

\newcommand{\macheps}{\varepsilon_{\mathrm{m}}}
\newcommand{\sgn}{\mathrm{sgn}}

\begin{title}
{\LARGE {\bf A Numerical Algorithm for Zero Counting. \\
I: Complexity and Accuracy}}
\end{title}
\author{Felipe Cucker
\thanks{Partially supported by City University
SRG grant 7002106.}\\
Dept. of Mathematics\\
City University of Hong Kong\\
HONG KONG\\
e-mail: {\tt macucker@cityu.edu.hk}
\and
Teresa Krick
\thanks{Partially supported by grants UBACyT X112/06-09,
CONICET PIP 2461/00 and ANPCyT 33671/05.}\\
Departamento de Matem\'atica\\
Univ. de Buenos Aires \&\ CONICET\\
ARGENTINA\\
e-mail: {\tt krick@dm.uba.ar}
\and
Gregorio Malajovich\thanks{Partially supported by
CNPq grants 304504/2004-1, 472486/2004-7, 470031/2007-7,
303565/2007-1, and by FAPERJ grant E26/170.734/2004.
}\\
Depto. de Matem\'atica Aplicada\\
Univ. Federal do Rio de Janeiro\\
BRASIL\\
e-mail: {\tt gregorio@ufrj.br}
\and
Mario Wschebor\\
Centro de Matem\'{a}tica\\
Universidad de la Rep\'{u}blica\\
URUGUAY\\
e-mail: {\tt wschebor@cmat.edu.uy}
}

\date{}
\makeatletter
\maketitle
\makeatother

\begin{quote}
{\small
{\bf Abstract.} We describe an algorithm to count the number of distinct
real zeros of a polynomial (square) system $f$. The algorithm performs
$\Oh(\log(n\bD\kappa(f)))$ iterations (grid refinements) 
where $n$ is the number of polynomials (as well as the
dimension of the ambient space), $\bD$ is a bound on the polynomials'
degree, and $\kappa(f)$ is a condition number for the system. Each
iteration uses an exponential number of operations. The algorithm uses
finite-precision arithmetic and a major feature in our results is a 
bound for the precision required to ensure the returned output is 
correct which is polynomial in $n$ and $\bD$ and logarithmic in 
$\kappa(f)$. The algorithm parallelizes well in the sense that
each iteration can be computed in parallel time polynomial 
in $n$, $\log\bD$ and $\log(\kappa(f))$. 
}\end{quote}

\section{Introduction}

In recent years considerable attention was put on
the complexity of counting problems over the reals. The counting
complexity class $\CPRi$ was introduced~\cite{meer:00} and
completeness results for $\CPRi$ were established~\cite{BC03} for
natural geometric problems notably, for the computation
of the Euler characteristic of semialgebraic sets.
As one could expect, the ``basic'' $\CPRi$-complete problem
consists of counting the real zeros of a system of polynomial
equations.

Algorithms for counting real zeros have existed since long. One such
algorithm follows from the work of Tarski~\cite{Tarski51} on
quantifier elimination for the theory of the reals. Its complexity
is hyperexponential. Algorithms with improved complexity (doubly
exponential) were devised in the 70s by Collins~\cite{Collins} and
W\"utrich~\cite{Wut76}. A breakthrough was reached a decade later
with the introduction of the critical points method by Grigoriev and
Vorobjov~\cite{GriVo88,Gri88} which uses exponential time.
Algorithms counting connected components (and hence, in the
zero-dimensional case, solutions) based on this method can be found
in~\cite{GriVo92,HeRoSo:94}, and in the straight-line program model
of computation in~\cite{BaGiHePa:05}. These algorithms parallelise
well in the sense that one can devise versions of them working in
parallel polynomial time when an exponential number of processors is
available. The $\CPRi$-completeness of the problem strongly
indicates that this is the best we can hope for.

All the algorithms mentioned above are ``symbolic algorithms.''
They have been devised upon the premise that no perturbation or
round-off error is present. Were this not the case, it is not
difficult to see that errors would accumulate quite badly. Roughly
speaking, these algorithms construct some object of exponential
size  on which some basic computation (e.g., linear algebra)
is eventually performed. A question is posed, can one devise
``numerical algorithms'' (maybe iterative, which need not
terminate for ill-posed inputs) with a better behavior viz
the accumulation of round-off errors? For the problem of deciding
the existence of (or computing) a zero of a polynomial system
such algorithms were given in~\cite{CS98,Cucker99b,Li03}.
The goal of this article is to describe and analyze a
numerical algorithm for zero counting. We will do so by
developping appropriate versions of the tools used
in~\cite{CS98,Cucker99b}.
\smallskip

Let $d_1,\ldots,d_n\in\N$ and $\mathbf d = (d_1, \dots, d_n)$. We
will denote by $\Hd$ the space of polynomial systems
$f=(f_1,\ldots,f_n)$ with $f_i\in\R[X_0,\ldots,X_n]$ homogeneous of
degree $d_i$.

Zero rays of polynomial systems $f \in \Hd$ are associated
to pairs of zeros $(-\zeta,\zeta)$ of the restriction
$f_{|S^n}$ of
$f$ to the $n$-dimensional unit sphere $S^n\subset\R^{n+1}$.
Thus, it will be convenient
to consider a system $f \in \Hd$ as a (central symmetric,
analytic) mapping of $S^n$ into $\mathbb R^n$. If we denote by
$Z(f) = \{\zeta\in S^{n} : f(\zeta)=0\}$ the zero-set of $f$
in $S^n$ then the number $\nrr(f)$ of zero rays of the system
$f$ is half the cardinality of $Z(f)$.

In this paper we describe a finite-precision algorithm computing $\nrr(f)$,
given $f\in\Hd$. To analyze its complexity and accuracy, besides the
number $n$ of polynomials, we will rely on two more additional
parameters. One is $\bD=\max_{i\leq n} d_i$. The other is a
condition measure $\kappa(f)$ for the system $f$. We will describe
this measure in detail in Section~\ref{sec:basics} below. We will
also let $S=\max S_i$ where $S_i$ is the number of non-zero coefficients
of $f_i$. Note that $S$ is bounded by a simple expression in terms of
$n$ and $\bD$, namely, $S={n+\bD\choose \bD}$. Yet, we will express
dependancy on $S$
since this may be relevant for the case of sparse systems of
polynomials. Our main result is the following.

\begin{theorem}\label{th:main}
There exists an iterative algorithm which, with input $f\in\Hd$,
\begin{description}
\item[(1)]
Returns $\nrr(f)$.

\item[(2)]
Performs $\Oh(\log (n\bD \kappa(f)))$
iterations and has a total cost (number of arithmetic
operations) of
$$
  \Oh\left(\log (n\bD \kappa(f))
    (n+1)^2\left(\frac{2(n+1)\bD^2\kappa(f)^2}{\a_*}\right)^{2n}
    \right),
$$
where $\a_*\approx 0.0384629388\ldots$ is a universal constant.
\item[(3)]
Can be well-parallelized in the sense that it admits a
parallel version running in time
$$
  \Oh(n^2\ln(n\bD\kappa(f))(\ln(n\bD\kappa(f))^2+\ln(\a_*)^2))
$$
with a number of processors exponential in this quantity.
\item[(4)]
Can be implemented with finite precision (both versions,
sequential and parallel). The running time remains the
same (with $\a_*$ replaced by $\ab\approx 0.028268\cdots$)
and the returned value is $\#_{\R}(f)$ as long as the machine
precision (i.e., the round-off unit) $u$ satisfies
$$
 u\leq\frac{1}{\Oh\left(\bD^2n^{5/2}\kappa(f)^3(\log S+n^{3/2}
\bD^2\kappa(f)^2)\right)}.
$$
\item[(5)]
It can be modified to return, in addition and 
for each real zero $\zeta\in S^n$ of $f$, an approximate zero 
$x$ of $f$ in the sense that Newton's iteration, starting at $x$, 
converges to $\zeta$ quadratically fast.
\end{description}
\end{theorem}

\begin{remark}
{\bf (i)\quad }
A system $f$ for which arbitrarily small perturbations
may change the value $\nrr(f)$ is considered {\em ill-posed}
in our context since for arbitrarily small machine precisions
finite precision algorithms may return an incorrect value.
Consequently, the condition number $\kappa(f)$ is infinite
in these cases (and only then). This happens when
$f$ has multiple real zeros and, in particular, when $f$
has infinitely many real zeros. In these cases the algorithm
of Theorem~\ref{th:main} may not halt.

\noindent
{\bf (ii)\quad} Numerical algorithms compute functions $\varphi$ 
on real data. Error analysis for algorithms computing (vectors of) 
real numbers ---i.e. for which the image of $\varphi$ has 
non-empty interior--- are usually expressed in terms of bounds 
for the relative error of the computed quantities. That is, 
for data $d$, bounds in 
$$
     \frac{\|\varphi(d)-\fl(\varphi(d)\|}{\|\varphi(d)\|}
$$
where $\fl(\varphi(d))$ is the vector actually computed with 
finite precision. This relative error varies 
continuously with $d$ and depends on the condition 
of $d$ and on the precision $u$. Such a form of analysis, however, 
becomes meaningless when computing quantities taking a finite 
number of values. Indeed, if $R_a$ denotes the set of input data 
$d$ for which $\varphi(d)=a$ the following happens. When $d$ is 
in the interior of $R_a$ we have that the relative error above 
is 0 for sufficiently small $u$. In contrast, when $d$ is on the 
boundary of $R_a$, that error may remain constant for all $u>0$. 
Because of this, error analysis for this kind of discrete-valued 
problems has a different form, as in Theorem~\ref{th:main}. One bounds 
how small $u$ needs to be to guarantee a correct answer. Such a bound, 
needless to say, also depends on the condition of the data $d$. Examples 
of this type of analysis can be found 
in~\cite{ChC03,Cucker99b,CP01,CS98}. In each of these references a 
condition number for the problem at hand occurs in the error analysis. 
We note that the one in~\cite{Cucker99b} is essentially our 
$\kappa(f)$.    
\end{remark}

The rest of the paper is organized as follows. In 
Section~\ref{sec:basics} we describe the basic objects we 
will deal with as well as fixing the notation. In 
Sections~\ref{sec:exc} and~\ref{sec:newton} we prove the two 
technical results our algorithm relies on. In 
Section\ref{sec:ip} we describe the algorithm under the 
assumption of infinite precision and we prove parts (1), (2), 
and (3) of Theorem~\ref{th:main}. The geometric ideas making the 
algorithm work are best seen in this context. 
Section~\ref{sec:fp} then describes the necessary modifications 
to make the algorithm work as well under finite precision. These 
modifications are simple and can be summarized by saying that we 
relax a bit the inequalities tested in the algorithms to make 
room for the finite-precision errors to fit in. 

\section{Preliminaries}\label{sec:basics}

Denote by $\HH_d$ the subspace of $\R[X_0,\ldots,X_n]$ of
homogeneous polynomials of degree $d$. Then,
$\Hd=\HH_{d_1}\times \cdots\times \HH_{d_n}$.

If $g\in \HH_d$ we write
\[
  g(X) = \sum_{J} g_J X^J
\]
where $J=(J_0, \dots, J_n)$ is assumed to range over all
multi-indices such that $|J| = \sum_{k=0}^n J_k = d$,
$X^J = X_0^{J_0} X_1^{J_1}\dots X_n^{J_n}$ and $g_J\in\R$.
Multinomial coefficients are defined by:
\[
  {d \choose J} = \frac{d!}{J_0 ! J_1 ! \cdots J_n !} .
\]

The space $\HH_d$ is endowed with the inner product
$$
   \langle g,h \rangle =
   \sum_{|J|=d}
   \frac{ g_J h_J }{{d \choose J}}
$$
which gives rise to the norm $\|g\|=\sqrt{\langle g,g\rangle}$.
These norms, for
$d_1,\ldots,d_n$, induce a norm in $\Hd$ by taking for
$f=(f_1,\dots,f_n)\in \Hd$:
$$
  \|f\|= \|(f_1,\ldots,f_n)\|=\max_{1\le i\leq n}\|f_i\|.
$$
Let $O(n+1)$ be the orthogonal group. The inner product above is
known to be $O(n+1)$-invariant: for all $Q\in O(n+1)$ and all
$g,h\in\HH_ d$,
\[
 \langle g \circ Q, h \circ Q \rangle = \langle g , h \rangle .
\]
(This is a direct consequence of \cite[III-7]{Weyl32} or
\cite[Theorem 1 p. 218]{BCSS98}, by considering $O(n+1)$ as subgroup
of $U(n+1)$). The associated norm $\|f\|$ on $\Hd$ is therefore also
$O(n+1)$-invariant. We will use this norm on $\Hd$ all along this
paper. For $x=(x_1,\ldots,x_n)\in\R^n$ we recall that
$\|x\|_2=(x_1^2+\cdots+x_n^2)^{1/2}$ and
$\|x\|_\infty=\max\{|x_1|,\ldots,|x_n|\}$. We will often denote
$\|x\|_2$ simply by $\|x\|$.

For $f\in\HH_{\bd}$ and $x\in S^{n}$ define
\begin{equation}\label{eq:mu}
  \mu_{\rm norm}(f,x) = \|f\|\sqrt{n} \left\| Df(x)_{|T_x S^n}^{-1}
  \left[
  \begin{matrix} \sqrt{d_1} \\ & \sqrt{d_2} \\
   & & \ddots \\ & & & \sqrt{d_n}
  \end{matrix} \right] \right\|
\end{equation}
where $Df(x)_{|T_x S^n}$ is the restriction to the tangent space of
$x$ at $S^n$ of the derivative of $f$ at $x$ and the norm is the
spectral norm, i.e. the operator norm with respect to $\|\ \|_2$. We
now define the {\em condition number} $\kappa(f)$ of $f\in \Hd$:
$$
   \kappa(f)=\max_{x\in S^n} \min
         \left\{\mu_{\rm norm}(f,x),\frac{\|f\|}{\|f(x)\|_\infty}\right\}.
$$

\begin{remark}
The quantity $\kappa(f)$ is closely related to other condition
numbers for similar problems.

A version of the quantity $\mu_{\rm norm}(f,\zeta)$ was
introduced in~\cite{Bez1,Bez3,Bez4}
(see also~\cite[Chapter~12]{BCSS98}) for a complex polynomial
system $f$ and a zero $\zeta$ of $f$ in the complex unit sphere
$S_{\C}^n\subset\C^{n+1}$.
The {\em normalized condition number} of such a system
$f$ was then defined to be
\begin{equation}\label{muC}
  \mu_{\rm norm}(f):=\max_{\zeta\in S^n_{\C}\mid f(\zeta)=0}
  \mu_{\rm norm}(f,\zeta).
\end{equation}
Actually, the version of $\mu_{\rm norm}(f,\zeta)$
introduced in~\cite{Bez1,Bez3,Bez4} differs from
(\ref{eq:mu}) in the fact that $\|f\|$ is defined
as $(\sum \|f_i\|^2)^{1/2}$ (and there is no $\sqrt{n}$
factor). It is bounded above by the expression in~(\ref{eq:mu}).

Over the reals, the right-hand side in~(\ref{muC})
may not be well-defined since
the zero set of $f$ may be empty. In~\cite{CS98} real systems were
considered (as in the present paper) and
an algorithm deciding feasibility of $f$ (i.e., whether $f$ has a
real zero) was proposed. Its complexity was analyzed in terms of a
condition number which, using our notation and
modulo minor details, is defined as follows
$$
  \left\{\begin{array}{ll}
   \displaystyle\min_{\zeta\in S^n\mid f(\zeta)=0} \mu_{\rm norm}(f,\zeta)
   & \mbox{if $f$ is feasible}\\[8pt]
   \displaystyle\max_{\zeta\in S^n} \frac{\|f\|}{\|f(\zeta)\|_\infty}
   & \mbox{if $f$ is infeasible.}
  \end{array}\right.
$$
Note the use of $\min$ (instead of $\max$) in the first line above.
This is due to the fact that the time needed for the algorithm
in~\cite{CS98} to detect the existence of a zero depends on
the best conditioned zero of $f$. The existence of other,
poorly conditioned (or even singular), zeros of $f$ is
irrelevant.

Shortly after, the algorithm in~\cite{CS98} was
extended to an algorithm which would, in addition and if $f$ is
feasible, return a zero of $f$~\cite{Cucker99b}.
The complexity of this extension was
studied in terms of a condition number (denoted $\varrho(f)$
in~\cite{Cucker99b}) which,
essentially, coincides with our $\kappa(f)$.
\end{remark}

\begin{proposition}\label{prop:kappa1}
For all $f\in\Hd$, $\kappa(f)\geq 1$.
\end{proposition}

\proof
Let $x\in S^n$. Because of orthogonal invariance,
we may assume without loss of generality that
$x=e_0:=(1,0,\ldots,0)$.

It is then immediate that $\|f(x)\|_\infty\leq \|f\|$. This shows
that the second expression in the definition of $\kappa$ is at least
1.

For the first expression, i.e., $\mu_{\rm norm}(f,x)$,
define $g=(g_1,\dots,g_n)\in\Hd$ by $g_i(X) = f_i(X) - f_i(e_0)
X_0^{d_i}$. Then $g(e_0)=0$ and~\cite[Corollary 3 p. 234]{BCSS98},
$\mu_{\rm norm}(g,e_0)\geq 1$ (this is shown for the version of
$\mu_{\rm norm}$ with the 2-norm for $\|f\|$, which is bounded above
by the expression~(\ref{eq:mu})). Since $Df(e_0)=Dg(e_0)$ and
$\|g\|\leq \|f\|$, we can conclude $\mu_{\rm norm}(f,e_0)\geq
\mu_{\rm norm}(g,e_0)\geq 1$. \eproof

\section{The exclusion Lemma}\label{sec:exc}

In this article,  $d(\ ,\ )$ denotes the Riemannian (angular)
distance in $S^n$ (which satisfies $0\le d(x,y)\le \pi, \ \forall \,
x,y\in S^n$) and for $x\in S^{n},r>0$, we set $B(x,r):=\left\{ y\in
S^{n}: d(y,x)<r\right\}$ and $\bar B(x,r):=\left\{ y\in S^{n}:
d(y,x)\le r\right\}$.

\smallskip
The following result can be used to support an exclusion test.

\begin{lemma}\label{exclusionlemma}
Let $f \in \Hd$ and let $x, y \in S^n$ such that $d(x,y)\le \sqrt
2$. Then,
\[
  \|f(x)-f(y)\|_\infty \le \|f\| \sqrt{\bD} \ d(x,y)
\]
In particular, if $f(x) \ne 0$, there is no zero of $f$ in
$B(x,\min\{ \|f(x)\|_\infty / (\|f\| \sqrt{\bD}), \sqrt 2\})$.
\end{lemma}

\proof
An immediate consequence of the definition
of the $O(n+1)$-invariant inner product is that $\HH_d$
endowed with this inner product is
a reproducing kernel Hilbert space~\cite[Prop.~2.21]{CZ:06}.
This implies that, for all $g\in\HH_d$ and $x\in\R^{n+1}$,
\begin{equation} \label{repkernel}
  g(x) = \langle g(X), (x^T X)^{\deg g} \rangle .
\end{equation}
Because of orthogonal invariance, we can assume that $x = \mathrm
e_0$ and $y = \mathrm e_0 \cos \theta + \mathrm e_1 \sin \theta$,
where $\theta = d(x,y)$. Equation~(\ref{repkernel}) implies that
\begin{eqnarray*}
   f_i(x)-f_i(y)
&=& \langle f_i(X), (x^T X)^{d_i} \rangle
    - \langle f_i(X), (y^T X)^{d_i} \rangle
    = \langle f_i(X), (x^T X)^{d_i} - (y^T X)^{d_i} \rangle \\
&=& \langle f_i(X), X_0^{d_i} - (X_0 \cos \theta
  + X_1 \sin \theta)^{d_i} \rangle.
\end{eqnarray*}
Hence, Cauchy-Schwarz-Bunyakowsky implies:
$$
   | f_i(x) - f_i(y) | \le \| f_i \|\,
   \| X_0^{d_i} - (X_0 \cos \theta + X_1 \sin \theta)^{d_i}\|.
$$
Since
$$
  X_0^{d_i} - (X_0 \cos \theta + X_1 \sin \theta)^{d_i}
  = X_0^{d_i} (1-(\cos \theta)^{d_i}) + \sum_{k=1}^{d_i}
 {d_i\choose k}(\cos \theta)^{d_i-k} (\sin \theta)^kX_0^{d_i-k}X_1^k,
$$
we have:

\begin{eqnarray}
\| X_0^{d_i} - (X_0 \cos \theta + X_1 \sin \theta)^{d_i} \|^2 &=&
(1- (\cos \theta)^{d_i}) ^2 + \sum_{k=1}^{d_i} {d_i\choose k}(\cos
\theta)^{2(d_i-k)} (\sin \theta)^{2k} \nonumber
\\
&=& (1- (\cos \theta)^{d_i}) ^2 +1 - (\cos \theta)^{2d_i} \nonumber
\\
&=& 2 (1 - (\cos \theta)^{d_i}) \nonumber
\\
&\le& 2 (1 - (1-\frac{\theta^2}{2})^{d_i}) \label{cos}\\
&\le& 2 (1 - (1-d_i \frac{\theta^2}{2})) \label{fcion} \\
&\le& d_i \theta^2, \nonumber
\end{eqnarray}
where the inequality in line (\ref{cos}) is obtained from Taylor
expanding $\cos\theta$ around 0, and the inequality in line
(\ref{fcion}) is due to the fact that $(1-a)^d\ge 1-da$ for $a\le
1$.
\\
We conclude that
\[
   | f_i(x) - f_i(y) | \le \| f_i \| \,\theta \, \sqrt{d_i}
\]
and hence
\[
   \| f(x) - f(y) \|_\infty \le \|f\| \, \theta \, \sqrt{\max_i d_i}.
\]
For the second assertion, we have
\begin{eqnarray*}\|f(y)\|_\infty
&\ge&  \|f(x)\|_\infty - \|f(x)-f(y)\|_\infty \\
&\ge & \|f(x)\|_\infty -  \|f\|\sqrt{\bD} \,d(x,y) \qquad \mbox{since
}
d(x,y)\le \sqrt 2 \\
&> & \|f(x)\|_\infty -  \|f\|\sqrt{\bD}
\,\|f(x)\|_\infty/(\|f\|\sqrt{\bD}) \quad = \quad 0.
\end{eqnarray*}
\eproof

\section{The proximity Theorem}\label{sec:newton}

\subsection{Newton and Smale}

Newton iteration on the sphere $S^n$ is defined by
\[
\begin{array}{rrcl}
 N_f: & S^n & \rightarrow & S^n \\
        & x & \mapsto & N_f(x)
    = \exp_x \left(-Df(x)_{|T_x S^n}^{-1} f(x) \right)
\end{array}
\]
where $\exp_x$ is the exponential map at $x$,
\[
\exp_x h = \cos(\|h\|) x + \frac{\sin(\|h\|)}{\|h\|} h .
\]

Furthermore, the standard invariants of $\alpha$-theory,
introduced by Smale in~\cite{Smale86},
can be defined as:
\begin{eqnarray*}
\beta(f,x) &=&
\left\|Df(x)_{|T_x S^n}^{-1} f(x)\right\| , \\
\gamma(f,x) &=&
\sup_{k \ge 2} \left\|
\frac{ Df(x)_{|T_x S^n}^{-1} D^kf(x)_{|(T_x S^n)^k}}{k!}
\right\|^{1/(k-1)} ,\\[1mm]
\alpha(f,x) &= &\beta(f,x) \gamma(f,x).
\end{eqnarray*}

\begin{remark}\label{rem:beta} {\ }
\begin{description}
\item[(i)]
It is easy to see that $\beta(f,x)=d(x,N_f(x))$.
\item[(ii)]
We will not use Newton's method in our algorithm. We are instead
interested in its alpha theory which guarantees existence of zeros
near points $x$ with $\alpha(f,x)$ small enough.
\item[(iii)]
The Newton iteration presented above is not the iteration known as
`projective Newton'. There is an alpha theory for that method,
available in~\cite{Malajovich94}.
\end{description}
\end{remark}

Here we use slight modifications of the quantities $\alpha, \beta$
and $\gamma$, more adapted to our purposes. We set
\begin{eqnarray*}\bar\beta(f,x)&:= &\mu_{\rm
norm}(f,x)\frac{\|f(x)\|_\infty}{\|f\|} \\
\bar\gamma(f,x)&:= & \frac{\bD^{3/2}}{2}\mu_{\rm norm}(f,x)\\[1mm]
\bar\alpha(f,x)&:=&\bar\beta(f,x)\bar\gamma(f,x).\end{eqnarray*}

The definition of $\bar \gamma$ is motivated by  the  estimate of
$\gamma$~\cite[Theorem 2 p. 267]{BCSS98}.

$$
\gamma(f,x) \le \bar\gamma (f,x).$$ which  yields the lower bound
\begin{equation}\label{eq:lower_kappa}
 \kappa(f)\geq \max_{\zeta \mid f(\xi)=0} 2\bD^{-3/2}\gamma(f,\zeta).
\end{equation}

We also observe that $\bar \gamma(f,x)\ge \frac{\bD^{3/2}}{2}$ since
$\mu_{\rm norm}(f,x)\ge 1$ and that $\beta(f,x)\le \bar\beta(f,x)$
since
$$\beta(f,x)=
\left\|Df(x)_{|T_x S^n}^{-1} f(x)\right\|\le \sqrt n \|f(x)\|_\infty
\left\|Df(x)_{|T_x S^n}^{-1} \right\|\le  \mu_{\rm norm}(f,x)
\frac{\|f(x)\|_\infty
     }{\|f\|}=\bar\beta(f,x).
$$

Therefore $\alpha(f,x)\le \bar \alpha(f,x)$.

%***This, in turn, yields a lower bound for $\kappa(f)$,
%\begin{equation}\label{eq:lower_kappa}
% \kappa(f)\geq \max_{\zeta \mid f(\xi)=0} 2\bD^{-3/2}\gamma(f,\zeta).
%\end{equation}***

\subsection{Proximity and unicity from data at a point}

\begin{definition}
We say that $x \in S^n$ is an {\em approximate zero} for $f$ if and
only if the Newton sequence $\{x_k\}_{k \in \mathbb N}$, where
$x_0:=x$ and  $x_{k+1} := N_{f}(x_k)$, is defined for all $k$ and
moreover
\[
  d(x_k,x_{k+1}) \le
 \left( \frac{1}{2} \right)^{2^k -1} d(x_0,x_1) .
\]
The limit point $\zeta = \lim_{k \rightarrow \infty} x_k$ is a fixed
point for Newton iteration and a zero of $f$. It is called the {\em
associated zero} to $x$.
\end{definition}

In what follows we denote  $\sigma: = \sum_{k \ge 0} 2^{-2^k + 1} =
1.632843018\dots$ and we set $$ \bar B_f(x)  :=  \{y\in S^n \mid
d(x,y)\le \sigma\bar \beta(f,x)\}.$$

The main technical tool in our algorithm is provided by the
following result.

\begin{theorem}\label{cor:alpha}
There exists an universal constant  $\a_*:=0.0384629388\dots$  such
that for all $x\in S^n$, if \ $
 \bar\alpha(f,x)<
 \a_*$, \
then
\begin{description}
\item[(i)]
$x$ is an approximate zero of $f$.
\item[(ii)]
If $\zeta$ denotes its associated zero then $\zeta\in \bar B_f(x)$.
\item[(iii)]
Furthermore, for each point $z$ in $\bar B_f(x)$ the Newton sequence
starting at $z$ converges to $\zeta$.
\end{description}
\end{theorem}

\subsection{Background material}

Theorem~\ref{cor:alpha} is a consequence of the following two
results, which are restatements of results proved in~\cite{DPM}.
While~\cite{DPM} deals with Newton iteration on arbitrary complete
real analytic Riemannian manifolds, here we  reword them in terms of
Newton iteration on the unit sphere $S^n$ (Example $1$ in
\cite{DPM}). The $\gamma$-Theorem for mappings~\cite[Theorem
1.3]{DPM} becomes the following.

\begin{theorem}\label{cor:1}
Let $f: S^n\to\R^n$ be analytic. Suppose that
$f(\zeta)=0$ and $Df(\zeta)$ is an isomorphism. Let
\[
R(f,\zeta) := \min \left\{ \pi , \frac{3 - \sqrt{7}}{2
\gamma(f,\zeta)} \right\} \ .
\]
If $d(x,\zeta) \le R(f, \zeta)$, then the Newton sequence
$x_k = N_f^k (x)$ is defined for all $k \ge 0$ and
$d(x_k, \zeta) \le \left( \frac{1}{2} \right)^{2^k -1} d(x,\zeta)$.
In particular, $\{x_k\}$ converges to $\zeta$.
\end{theorem}

\smallskip
Now let $\alpha_0 := 0.130716944\dots$ denote the smallest positive
root of the polynomial $\psi(u)^2 - 2u$, and
\[
  s_0 := \frac{1}{\sigma + \frac{(1-\sigma \alpha_0)^2}
 {\psi(\sigma \alpha_0)}
 \left(1+\frac{\sigma}{1-\sigma \alpha_0}\right)}
 =  0.103621842\dots
\]
We  state  the $\alpha$-Theorem for mappings~\cite[Theorem 1.4]{DPM}
for the sphere $S^n$.

\begin{theorem}\label{cor:2}
Let $f: S^n\to\R^n$ be analytic. Let $x \in  S^n$
be such that $\beta(f,x) \le s_0 \pi$ and $\alpha(f,x) \le \alpha_0$.
Then the Newton sequence
$x_k = N_f^k (x)$ is defined for all $k \ge 0$ and
converges to a zero $\zeta$ of $f$. Moreover,
\[
d(x_k,x_{k+1}) \le \left( \frac{1}{2} \right)^{2^k -1} \beta(f,x)
\]
and
\[
d(x_k, \zeta) \le \sigma \beta(f,x).
\]
\end{theorem}

\smallskip
\noindent Finally  we introduce  $\psi(u) := 1 - 4u + 2 u^2$, which
is positive and decreasing for $0<u<1 - \frac{\sqrt{2}}{2}$, and
state~\cite[Lemma 4.3]{DPM}:

\begin{lemma}\label{gamma}
Let $x, y \in S^n$ with $d(x, y) < \pi$. Suppose that $Df(x)$ is
nonsingular and
\[
  \nu := d(x,y) \gamma(f,x) < 1 - \frac{\sqrt{2}}{2}.
\]
Then
\[
 \gamma(f,y) \le \frac{ \gamma(f,x) }{ (1-\nu) \psi(\nu)}.
\]
\end{lemma}

\subsection{Proof of Theorem~\ref{cor:alpha}}

Set $\nu_*: = 0.0628039411\dots$ for the only real root of the
polynomial
%\begin{equation}\label{nu}
  $$\Psi(u):=(3-\sqrt{7}) (1-u) \psi(u) - 4u,$$
%\end{equation}
%CALCULOS USANDO octave (compatible con matlab):
%
%           k=3-sqrt(7) ;
%           g=[-2,6,-5,1]*k-[0,0,2,0]
%           r=roots(g)
%           rr=r(3)
%           printf("%40.30f\n",rr) ;
%           0.099308571344556492710431427895
% Not all numbers need to be significative !
%   octave:37> polyval(g,rr+ 1e-14)
%   ans =  -3.3695e-14
%   octave:38> polyval(g,rr- 1e-14)
%   ans =  3.3751e-14
%
%
%            sigma=1; for k=1:100, sigma=sigma+2^(-2^k+1); end ;
% octave:43> printf("%40.30f\n",sigma) ;
%        1.632843018043786287307739257812
% octave:44> printf("%40.30f\n",rr/sigma) ;
%        0.060819423696671273682490266310
and $\alpha_*:=\frac{\nu_*}{\sigma}=0.0384629388\dots$. Note that
$\alpha_*\leq\min\{\alpha_0,s_0\pi\}$.
%We define
%\[
%\a_* = \min \left\{
%\alpha_0,
%\nu_*/\sigma,
%s_0 \pi,
%\right\}.
%\]
%Numerically, $\a_* = \nu_* / \sigma = 0.060819423\cdots$.

\smallskip

Since $\bar\gamma(f,x)\ge \frac{\bD^{3/2}}{2}$, the hypothesis of
Theorem~\ref{cor:2} hold from $\alpha(f,x)\le
\bar\alpha(f,x)<\alpha_*\le \alpha_0$ and $\beta(f,x)\le
\bar\beta(f,x)\le
\frac{2\bar\alpha(f,x)}{\bD^{3/2}}<\frac{2\a_*}{\bD^{3/2}}<s_0\pi$.
\\
Using Remark~\ref{rem:beta}(i) it follows that $x$ is an approximate
zero of $f$, and that the associated zero $\zeta$ satisfies:
\[
d(x,\zeta) \le \sigma \beta(f,x)\le \sigma \bar\beta (f,x).
\]
This already proves Parts~(i) and~(ii) of Theorem~\ref{cor:alpha}.
\\
We show (iii). Since $d(x,\zeta)\le \sigma\bar\beta(f,x)<\sigma
s_0\pi <\pi$,
$$ \nu = d(x,\zeta)\gamma(f,x) \le
\sigma\bar\beta(f,x)\gamma(f,x)\le
  \sigma \bar\alpha(f,x)\le
  \sigma \a_* =\nu_*< 1 - \frac{\sqrt{2}}{2},$$
and we can apply Lemma~\ref{gamma}. Therefore
$$
  4\sigma\bar\beta(f,x)\gamma(f,\zeta)
\leq 4\sigma\bar\beta(f,x)\gamma(f,x)\frac{1}{(1-\nu)\psi(\nu)} \leq
4\nu_*\frac{1}{(1-\nu_*)\psi(\nu_*)} = 3-\sqrt{7},$$ because
$(1-u)\psi(u)$ decreases for $0<u<1 - \frac{\sqrt{2}}{2}$, and
$\nu_*$ is a zero of $(3-\sqrt{7}) (1-u) \psi(u) - 4u$. This shows,
since $2\sigma \bar\beta(f,x)\le \pi$, that
%\begin{equation}\label{eq:radius}
 $$ 2\sigma\beta(f,x)\leq R(f,\zeta)=\min\left\{\pi,
    \frac{3-\sqrt{7}}{2\gamma(f,\zeta)}\right\}.$$
%\end{equation}
We conclude applying Theorem~\ref{cor:1} to $z\in  \bar B_f(x)$,
since
$$
  d(z,\zeta)\leq d(z,x)+d(x,\zeta)\leq 2\sigma\bar\beta(f,x)
  \leq R(f,\zeta).
$$
It follows that the Newton sequence $\{z_k\}_{k\in\N}$ starting at
$z$ converges to $\zeta$.

\begin{remark}
The hypothesis on the radius of injectivity in~\cite{DPM} was
recently found to be redundant.
\end{remark}

\section{Infinite precision}\label{sec:ip}

\subsection{Grids and Graphs}

Our algorithm works on a grid on $S^n$. We easily construct one by
projecting onto $S^n$ a grid on the cube $C^n=\{y\mid
\|y\|_\infty=1\}$. We make use of the (easy to compute) bijections
$\phi:C^n\to S^n$ and $\phi^{-1}:S^n\to C^n$ given by
$\phi(y)=\frac{y}{\|y\|}$ and $\phi^{-1}(x)=\frac{x}{\|x\|_\infty}$.

Given $\eta:=2^{-k}$ for some $k\geq 1$, we consider the uniform
grid $\UU_\eta$ of mesh $\eta$ on $C^n$. This is the set of points
in $C^n$ whose coordinates are of the form $i2^{-k}$ for
$i\in\{-2^k, -2^k+1,\ldots,2^k\}$, with at least one coordinate
equal to 1 or $-1$. We denote by $\GG_\eta$ its image
by $\phi$ in $S^n$. Note that, for $y_1,y_2\in C^n$,
\begin{equation}\label{eq:b1}
  d(\phi(y_1),\phi(y_2))\leq \frac{\pi}{2}\|y_1-y_2\|_2
  \leq\frac{\pi}{2}\sqrt{n+1}\,\|y_1-y_2\|_\infty.
\end{equation}

Given $\eta$ as above we associate to it a graph $G_\eta$ as
follows. We set $A(f):=\{x\in S^n \mid \bar\alpha(f,x)<\a_*\}$. The
vertices of the graph are the points in $\GG_\eta\cap A(f)$. Two
vertices $x,y\in\GG_\eta$ are joined by an edge if and only if $\bar
B_f(x)\cap \bar B_f(y)\neq\emptyset$.

Note that as a simple consequence of Theorem~\ref{cor:alpha} we
obtain the following lemma.

\begin{lemma}\label{lem:D(x)}
{\ }
\begin{description}
\item[(i)] For each $x\in A(f)$ there exists $\zeta_x \in Z(f)$
such that $\zeta_x \in \bar B_f(x)$. Moreover for each point $z$ in
$\bar B_f(x)$, the Newton sequence starting at $z$ converges to
$\zeta_x$.

\item[(ii)] Let $x,y\in A(f)$. Then
$\zeta_x=\zeta_y \iff  \bar B_f(x)\cap \bar B_f(y)\neq\emptyset$.
\eproof
\end{description}
\end{lemma}

We define  $Z(G_\eta):=\bigcup_{x\in G_\eta} \bar B_f(x) \subset
S^n$ where $x\in G_\eta$ has to be understood as $x$ running over
all the vertices of $G_\eta$. Similarly, for a connected component
$U$ of $G_\eta$, we define
$$
    Z(U):=\bigcup_{x\in U} \bar B_f(x).
$$

\begin{lemma}\label{lem:!}  {\ }
\begin{description}
\item[(i)]
 For each component $U$  of $G_\eta$, there is a unique
zero $\zeta_U\in Z(f)$ such that $\zeta_U\in Z(U)$. Moreover,
$\zeta_U \in \cap_{x\in U} \bar B_f(x)$.
\item[(ii)] If $U$ and $V$ are different components of $G_\eta$, then
$\zeta_U\ne \zeta_V$.
\end{description}
\end{lemma}

\proof (i) Let $x\in U$. Since $x\in A(f)$, by Lemma \ref{lem:D(x)}
(i) there exists a zero $\zeta_x$ of $f$ in $\bar B_f(x)\subseteq
Z(U)$. This shows the existence.  For the second assertion and the
uniqueness, assume that there exist $\zeta$ and $\xi$  zeros of $f$
in $Z(U)$. Let $x,y\in U$ be such that $\zeta\in \bar B_f(x)$, and
$\xi\in \bar B_f(y)$. Since $U$ is connected, there exist
$x_0=x,x_1,\ldots,x_{k-1},x_k:=y$ in $A(f)$ such that
$(x_i,x_{i+1})$ is an edge of $G_\eta$ for $i=0,\ldots,k-1$, that
is, $\bar B_f(x_i)\cap \bar B_f(x_{i+1})\neq\emptyset$. If $\zeta_i$
and $\zeta_{i+1}$ are the associated zeros of $x_i$ and $x_{i+1}$ in
$Z(f)$ respectively, then by Lemma \ref{lem:D(x)}(ii) we have
$\zeta_i=\zeta_{i+1}$, and thus $\zeta=\xi\in \bar B_f(x)\cap \bar
B_f(y)$.
\\
(ii) Assume $\zeta_U=\zeta_V\in \bar B_f(x)\cap \bar B_f(y) \subset
Z(U)\cap Z(V)$, then $x$ and $y$ are joined by an edge and belong to
the same connected component. \eproof

\subsection{The infinite precision algorithm}

{\small
\algo
\>\> {\tt Count\_Roots\_1}$(f)$\\
\>\> let $\eta:=\frac{2\sqrt{2}}{\pi\sqrt{n+1}}$\\
(1)  \>\> let $U_1,\ldots,U_r$ be the connected components of
     $G_\eta$\\
\>\> if \\
\>\>\> (i) for $1\leq i<j\leq r$ \\
\>\>\>\> for all $x_i\in U_i$ and all $x_j\in U_j$,
  $d(x_i,x_j)>\pi\eta\sqrt{n+1}$ \\
\>\> and \\
\>\>\> (ii) for all $x\in\GG_\eta\setminus A(f)$,
     $\|f(x)\|_\infty> \frac{\pi}{2}\eta\sqrt{(n+1)\bD}\|f\|$\\
\>\> then HALT and return $r/2$\\
\>\> else $\eta:=\eta/2$ \\
\>\>\> go to (1)
\falgo
}

\subsection{Proof of Theorem~\ref{th:main}(1--3)} \label{sec:proof1}

\noindent {\bf Proof of Part (1)}\quad This proof requires some
arguments of convexity. We can naturally define spherical convex
hulls for sets of points in $H^n$, an open half-sphere in $S^n$. If
$x_1,\ldots,x_q\in H^n$ we define
$$
  \SCH(x_1, \ldots, x_q):=
%  \Cone(\CH(x_1, \ldots, x_q)) \cap S^n
    \Cone(x_1, \ldots, x_q) \cap S^n
$$
where
%$\CH(x_1,\ldots,x_q)$ is the (Euclidean) convex hull
%of $\{x_1, \dots, x_q\}$ in $\R^{n+1}$ and $\Cone(U)$ is the
%pointed cone in $\R^{n+1}$ spanned by $U\subseteq\R^{n+1}$.
$\Cone(x_1, \dots, x_q)$ is the smallest convex cone with
vertex at the origin and containing the points $x_1,\ldots,x_q$.
%That is, we replace the convex hull by its (central) projection
%on the sphere.
Alternatively, we have,
$$
  \SCH(x_1, \dots, x_q)=\left\{
  \frac{\lambda_1 x_1+\cdots+\lambda_q x_q}
  {\|\lambda_1 x_1+\cdots+\lambda_q x_q\|} \mid
  \lambda_1,\ldots,\lambda_q \geq 0, \sum \lambda_i=1\right\}.
$$
We will use the following fact.

\begin{lemma}\label{lem:hull}
Let  $x_1,...,x_q\in H^{n}\subset \R^{n+1}$. If
$\bigcap_{i=1}^{q}\bar B(x_i,r_i)\neq \emptyset$, then
$\SCH(x_1,\ldots,x_q)\subset\bigcup_{i=1}^q\bar B(x_i,r_i)$.
\end{lemma}

\proof Let $x\in \SCH(x_1,\ldots,x_q)$ and $y\in
\bigcap_{i=1}^{q}\bar B(x_{i},r_{i})$. We will prove that $x\in \bar
B(x_{i},r_{i})$ for some $i$.

If $x=y$, this is obvious.

If $x\neq  y$, let \ $H$ be the half-space
\begin{equation*}
  H:=\left\{ z\in \R^{n+1}:\left\langle z,y-x\right\rangle < 0\right\}.
\end{equation*}

Since $\|x\|=\|y\|=1$, we have $\langle x+y,y-x\rangle =0$, and we
note that in this case, $x+y$ determines the mid-line between $x$
and $y$. Moreover, since $x\ne y$, we have $x\in H$ since $\langle
x,y-x\rangle = \langle x,y\rangle - \|x\|^2 < \|x\|\,\|y\| -\|x\|^2
=0$. Therefore the half-space $H$ is the set of points $z$ in
$\R^{n+1}$ such that the Euclidean distance $\|z-x\|< \|z-y\|$.

On the other hand, $H$ must contain at least one point of the set
$\left\{x_{1},...,x_{q}\right\}$ since if this were not the case,
the convex set $\Cone(\CH(x_1,\ldots,x_q))$ would be contained in
$\{z: \langle z,y-x\rangle \ge 0\}$, contradicting $x\in
\SCH(x_1,\ldots,x_q)$. Let, therefore, $x_{i}\in H$. It follows that
\begin{equation*}
  \left\| x-x_{i}\right\| < \left\| y-x_{i}\right\|
\end{equation*}
which implies
\begin{equation}\tag*{\qed}
  d(x,x_{i})< d(y,x_{i})\le r_{i}.
\end{equation}
\medskip

We can now proceed. Assume the algorithm halts, we want to show that
if $r$ equals the number of connected components of $G_\eta$, then
$\nrr(f)=\# Z(f)/2=r/2 $. We already know by Lemma \ref{lem:!} that
each connected component $U$ of $G_\eta$ determines uniquely a zero
$\zeta_U\in Z(f)$. Thus it is enough to prove that $Z(f)\subset
Z(G_\eta)$.\\ Assume
 that there is a zero $\zeta$ of $f$ in $S^n$ such that $\zeta$
is not in $Z(G_\eta)$. Let $B_\infty(\phi^{-1}(\zeta),\eta):= \{y\in
\UU_\eta\mid \|y-\phi^{-1}(\zeta)\|_\infty\leq \eta\}=
\{y_1,\ldots,y_q\}$, the set of all neighbors of $\phi^{-1}(\zeta)$
in $\UU_\eta$, and let $x_i=\phi(y_i)$, $i=1,\ldots,q$. Clearly,
$\phi^{-1}(\zeta)$ is in the cone spanned by $\{y_1,\ldots,y_q\}$
and hence $\zeta\in\SCH(x_1,\ldots,x_q)$.

We claim that there exists $j\leq q$ such that $x_j\not\in A(f)$.
Indeed, assume this is not the case. We consider two cases.
\smallskip

\noindent {\bf  (a)}\quad All the $x_i$ belong to the same connected
component $U$ of $G_\eta$. By Lemma~\ref{lem:!} there exists a
unique zero $\zeta_U\in S^n$ of $f$ in $Z(U)$ and $\zeta_U\in \cap_i
\bar B_f(x_i)$.  We may apply Lemma~\ref{lem:hull} to deduce that
\begin{equation*}\label{eq:51}
 \SCH(x_1, \dots, x_q) \subseteq
 \bigcup \bar B_f(x_i).
\end{equation*}
It follows that, for some $i\in\{1,\ldots,q\}$, $\zeta\in \bar
B_f(x_i)\subseteq Z(U)$, contradicting that $\zeta\not\in
Z(G_\eta)$.
\smallskip

\noindent
{\bf  (b)}\quad
There exist $\ell \neq s$
and $1\leq i<j\leq r$ such that $x_\ell\in U_i$ and
$x_s\in U_j$. Since condition (i) in the algorithm is
satisfied, $d(x_\ell,x_s)>\pi\eta\sqrt{n+1}$. But,
by \eqref{eq:b1},
$$
  d(x_\ell,x_s)\leq \frac{\pi}{2}\sqrt{n+1}\|y_\ell-y_s\|_\infty\leq
  \frac{\pi}{2}\sqrt{n+1}\left(\|y_\ell-\phi^{-1}(\zeta)\|_\infty
  +\|\phi^{-1}(\zeta)-y_s\|_\infty\right)
  \leq \pi\eta\sqrt{n+1},
$$
a contradiction.
\smallskip

We have thus proved the claim.
Let then $1\leq j\leq q$ be such that $x_j\not\in A(f)$.
Since condition~(ii) in the algorithm is satisfied
$\|f(x_j)\|_\infty> \frac{\pi}{2}\eta\sqrt{(n+1)\bD}\|f\|$. It follows
from the inequality $d(x_j,\zeta)\leq \frac{\pi}{2}\sqrt{n+1}\eta$
and Lemma~\ref{exclusionlemma} that $\|f(\zeta)\|_\infty>0$, a
contradiction.

\medskip

\noindent
{\bf Proof of Part~(2)}\quad
We need a few lemmas.

\begin{lemma}\label{lem:2roots}
If $\zeta_1\neq\zeta_2\in Z(f)$ then
$$
   d(\zeta_1,\zeta_2)\geq \frac{2(3-\sqrt{7})\bD^{-3/2}}{\kappa(f)}.
$$
\end{lemma}

\proof
For $i=1,2$, using (\ref{eq:lower_kappa})
and Proposition~\ref{prop:kappa1},
$$
  R(f,\zeta_i)=\min\left\{\pi,\frac{3-\sqrt{7}}{2\gamma(f,\zeta_i)}\right\}
  \geq \min\left\{\pi,\frac{(3-\sqrt{7})\bD^{-3/2}}{\kappa(f)}\right\}
  =\frac{(3-\sqrt{7})\bD^{-3/2}}{\kappa(f)}.
$$
Now suppose that $d(\zeta_1,\zeta_2)<R(f,\zeta_1)+R(f,\zeta_2)$ and
choose $x\in S^n$ such that $d(x,\zeta_1)<R(f,\zeta_1)$ and
$d(x,\zeta_2)<R(f,\zeta_2)$. Then Theorem~\ref{cor:1} implies that
$\zeta_1=\zeta_2$, a contradiction. \eproof

\begin{lemma}\label{lem:2points}
Let $x_1,x_2\in G_\eta$ with associated zeros $\zeta_1\ne \zeta_2$.
 If $\eta \le
\frac{2(3-\sqrt{7})\bD^{-3/2}}{3\pi\kappa(f)\sqrt{n+1}}$ then
$d(x_1,x_2)> \pi\eta\sqrt{n+1}$.
\end{lemma}

\proof Assume $d(x_1,x_2)\le\pi\eta\sqrt{n+1}$. Since $x_2\not\in
\bar B_f(x_1)$,  $d(x_1,x_2)> \sigma\bar\beta(f,x_1)$. Consequently,
$$
  d(x_1,\zeta_1)\le \sigma\bar\beta(f,x_1)< d(x_1,x_2)\le\pi\eta\sqrt{n+1}
$$
and, similarly, $d(x_2,\zeta_2)<\pi\eta\sqrt{n+1}$. But then,
$$
  d(\zeta_1,\zeta_2)\leq d(\zeta_1,x_1)+d(x_1,x_2)+d(x_2,\zeta_2)
  < 3\pi\eta\sqrt{n+1} \le
  \frac{2(3-\sqrt{7})\bD^{-3/2}}{\kappa(f)}
$$
contradicting Lemma~\ref{lem:2roots}.
\eproof

\begin{lemma}\label{lem:L2}
Let $x\in S^n$ such that $x\not\in A(f)$. If $\eta \le
\frac{\alpha_*}{(n+1)\bD^2\kappa(f)^2}$ then $\|f(x)\|_\infty
> \frac{\pi}{2}\eta\sqrt{(n+1)\bD}\|f\|$.
\end{lemma}

\proof Since $x\not\in A(f)$ we have $\bar\alpha(f,x)\geq \alpha_*$.
We divide the proof in two cases.
\medskip

\noindent
\fbox{{\bf Case I.}
$\min\left\{\mu_{\rm norm}(f,x),\frac{\|f\|}{\|f(x)\|_\infty}\right\}
=\frac{\|f\|}{\|f(x)\|_\infty}$}\medskip

\noindent
In this case
$$
  \eta\le \frac{\alpha_*}{(n+1)\bD^2\kappa(f)^2} \le
  \frac{\alpha_*\|f(x)\|_\infty^2}{(n+1)\bD^2\|f\|^2}
$$
which implies, since $\eta\leq\frac12<\frac{4\bD}{\pi^2\alpha_*}$,
$$
  \|f(x)\|_\infty\ge\frac{\sqrt{\eta} \sqrt{n+1}\bD\|f\|}{\sqrt{\alpha_*}}
  > \frac{\pi}{2}\eta\sqrt{(n+1)\bD}\|f\|.
$$
\medskip

\noindent
\fbox{{\bf Case II.}
$\min\left\{\mu_{\rm norm}(f,x),\frac{\|f\|}{\|f(x)\|_\infty}\right\}
=\mu_{\rm norm}(f,x)$}\medskip

\noindent
In this case
$$
  \eta\le \frac{\alpha_*}{(n+1)\bD^2\kappa(f)^2} \leq
  \frac{\alpha_*}{(n+1)\bD^2\mu_{\rm norm}(f,x)^2}
$$
which implies $\alpha_*\ge \eta(n+1)\bD^2\mu_{\rm norm}(f,x)^2$.
Also,
$$
  \alpha_*\leq
  \bar\alpha(f,x)=\frac12 \bar\beta(f,x)\mu_{\rm norm}(f,x) \bD^{3/2}
  \leq
      \frac1{2\|f\|} \mu_{\rm norm}(f,x)^2 \bD^{3/2}\|f(x)\|_\infty.
$$
Putting both inequalities together we obtain
$$
  \eta(n+1)\bD^2\mu_{\rm norm}(f,x)^2\le
  \frac1{2\|f\|} \mu_{\rm norm}(f,x)^2 \bD^{3/2}\|f(x)\|_\infty
$$
or yet,
\begin{equation}\tag*{\qed}
 \|f(x)\|_\infty\ge
  2\eta(n+1)\bD^{1/2}\|f\|> \frac{\pi}{2}\eta\sqrt{(n+1)\bD}\|f\|.
\end{equation}
\medskip

We can now conclude the proof of Part (2). Assume $\eta \le
\frac{\a_*}{(n+1)\bD^2\kappa(f)^2}$. Then the hypotheses of
Lemmas~\ref{lem:2points} and~\ref{lem:L2} hold. The first of these
lemmas ensures that condition~(i) in the algorithm is satisfied. The
second, that condition~(ii) is so. Therefore, the algorithm halts as
soon as $\frac{\a_*}{2(n+1)\bD^2\kappa(f)^2}< \eta \le
\frac{\a_*}{(n+1)\bD^2\kappa(f)^2}$. This gives a bound of
$\Oh(\ln(n\bD\kappa(f)))$ for the number of iterations. Since the
number of grid points considered at this iteration
($\eta=\frac{\a_*}{(n+1)\bD^2\kappa(f)^2}$) is at most
$2(n+1)\left(\frac{2(n+1)\bD^2\kappa(f)^2}{\a_*}\right)^{n}$, the
bound for the total complexity follows.

\medskip

\noindent
{\bf Proof of Parts~(3) and (5)}\quad
We have already seen that the number of iterations is bounded by
$\Oh(\ln(n\bD\kappa(f)))$. At each of these iterations, we
need to perform a number of computations on the (at most)
$2(n+1)\left(\frac{2(n+1)\bD^2\kappa(f)^2}{\a_*}\right)^{n}$
grid points to decide whether they are in $A(f)$. These can
be done independently. Then, we need to compute the number
of connected components of $G_\eta$. This can be
done (see, e.g.,~\cite{HaWa:90}) in
parallel time $\Oh(\ln(|V_\eta|))^2$ where $|V_\eta|$ denotes the
number of vertices of $G_\eta$ and therefore, in parallel time
at most $\Oh(n^2(\ln(n\bD\kappa(f))^2+\ln(\a_*)^2))$.
Since this is the dominant
step in the computation at a given iteration,
it follows that the total parallel time
consumed by the algorithm is at most
$\Oh(n^2\ln(n\bD\kappa(f))(\ln(n\bD\kappa(f))^2+\ln(\a_*)^2))$. 
This shows part~(3). For part~(5), just note that, for 
$i=1,\ldots,r$, any vertex $x_i$ of $U_i$ is an approximate zero 
of the only zero of $f$ in $Z(U_i)$.  
\eproof

\section{Finite Precision}\label{sec:fp}

\subsection{Making room to allow errors}

Our finite precision algorithm will be a variation of  Algorithm
{\tt Count\_Roots\_1}.  But since finite precision computations will
be affected by errors, we need to make room in the infinite
precision algorithm to allow them.  For this aim,  we state the
corresponding version of Theorem~\ref{cor:alpha}.

\begin{theorem}\label{th:alpha1}
There exist a universal constant $\ab=0.028268\cdots$ such that, for
all $x\in S^n$, if \ $
   \bar\alpha(f,x)<\ab
$, \ then
\begin{description}
\item[(i)]
$x$ is an approximate zero of $f$.
\item[(ii)]
If $\zeta$ denotes its associated zero then $\zeta\in \bar B_f(x)$.
\item[(iii)]
Furthermore, for each point $z$ s.t. $d(x,z)\le 2\sigma
\bar\beta(f,x)$ the Newton sequence starting at $z$ converges to
$\zeta$.
\end{description}
\end{theorem}

\proof Parts~(i) and~(ii) follow from Theorem~\ref{cor:alpha} and
the fact that $\ab<\a_*$. Part~(iii) is proved by taking
$\nu_{\bullet} = 0.046158\cdots$ to be the only real root of the
polynomial $\Psi(u):=(3-\sqrt{7}) (1-u) \psi(u) - 6u$, and
$\ab=\frac{\nu_{\bullet}}{\sigma}=0.028268$. Then, one proves as in
Theorem~\ref{cor:alpha} that $3\sigma\bar\beta(f,x)\leq
R(f,\zeta)$ from which it follows that, for all $z$ s.t. $d(x,z)\le
2\sigma\bar\beta(f,x)$,
$$
  d(z,\zeta)\leq d(z,x)+d(x,\zeta)\leq 3\sigma\bar\beta(f,x)
  \leq R(f,\zeta)
$$
and hence, that the Newton sequence
$\{z_k\}_{k\in\N}$ starting at $z$ converges to $\zeta$.
\eproof

The proofs of Lemmas~\ref{lem:2points} and~\ref{lem:L2} yield, {\em
mutatis mutandis}, the following results.

\begin{lemma}\label{lem:2pointsN}
Let $x_1,x_2\in G_\eta$ with associated zeros $\xi_1$ and $\xi_2$,
$\xi_1\neq\xi_2$. If $\eta \le
\frac{(3-\sqrt{7})\bD^{-3/2}}{3\pi\kappa(f)\sqrt{n+1}}$ then
$d(x_1,x_2) >2\pi\eta\sqrt{n+1}$.  \eproof
\end{lemma}

\begin{lemma}\label{lem:L2'}
Let $x\in S^n$ such that $\bar\alpha(f,x)>\frac{\ab}{3}$. If
$\eta\le\frac{\ab}{4\bD^2(n+1)\kappa(f)^2}$ then $\|f(x)\|_\infty >
\pi\eta\sqrt{(n+1)\bD}\|f\|$.\eproof
\end{lemma}

\subsection{Basic facts}\label{subsec:bf}

We recall the basics of a floating-point arithmetic which idealizes
the usual IEEE standard arithmetic. This system is defined by a set
$\F\subset\R$ containing $0$ (the {\em floating-point numbers}), a
transformation $r:\R\to\F$ (the {\em rounding map}), and a constant
$u\in\R$ (the {\em round-off unit}) satisfying $0<u<1$. The
properties we require for such a system are the following:
\begin{enumerate}
\item[(i)]
For any $x\in\F$, $r(x)=x$. In particular, $r(0)=0$.
\item[(ii)]
For any $x\in\R$, $r(x)=x(1+\d)$ with $|\d|\leq u$.
\end{enumerate}
We also define on $\F$ arithmetic operations following the classical
scheme
$$
  x\tilde\circ y=r(x\circ y)
$$
for any $x,y\in\F$ and $\circ\in\{+,-,\times,/\}$, so that
$$
  \tilde\circ:\F\x\F\to\F.
$$

The following is an immediate consequence of property (ii) above.

\begin{proposition}\label{prop:f1}
For any $x,y\in\F$ we have
\begin{equation}\tag*{\qed}
        x\tilde\circ y=(x\circ y)(1+\d),\qquad |\d|\leq u.
\end{equation}
\end{proposition}

When combining many operations in floating-point arithmetic,
quantities such as $\prod_{i=1}^n(1+\d_i)^{\rho_i}$ naturally
appear.
Our round-off analysis uses the notations and ideas
in Chapter~3 of~\cite{Higham96}, from where we quote
the following results:

%The proof of the following propositions can be found in
%Chapter~3 of~\cite{Higham96}.
%The notation they introduce, the
%quantities $\g_n$ and $\t_n$, and the relations showed therein, will
%be used in our round-off analysis.

\begin{proposition}\label{propHigham}
If $|\d_i|\leq u$, $\rho_i\in\{-1,1\}$, and $nu<1$, then
$$
  \prod_{i=1}^n(1+\d_i)^{\rho_i}=1+\theta_n,
$$
where
\begin{equation}\tag*{\qed}
        |\theta_n|\leq \g_n=\frac{nu}{1-nu}.
\end{equation}
\end{proposition}

\begin{proposition}\label{propHigham2}
For any positive integer $k$ such that $ku<1$, let $\theta_k, \theta_j$ be any
quantities satisfying
$$
        |\theta_k|\leq \g_k=\frac{ku}{1-ku}
\hspace{3em}
        |\theta_j|\leq \g_j=\frac{ju}{1-ju}
    .
$$
The following relations hold.
\begin{enumerate}
\item[{\rm 1.}]
$(1+\theta_k)(1+\theta_j)=1+\theta_{k+j}$ for some $|\theta_{k+j}|
\le \g_{k+j}$.

\item[{\rm 2.}]
$$
\frac{1+\theta_k}{1+\theta_j}=\left\{
\begin{array}{ll}
        1+\theta_{k+j}& \mbox{if $j\leq k$,}\\
        1+\theta_{k+2j}& \mbox{if $j> k$.}
\end{array}\right .
$$
for some $|\theta_{k+j}|
\le \g_{k+j}$ or some $|\theta_{k+2j}|
\le \g_{k+2j}$.
\item[{\rm 3.}]
If $ku,ju\leq1/2$, then $\g_k\g_j\leq \g_{\min\{k,j\}}$.

\item[{\rm 4.}]
$i\g_k\leq \g_{ik}$.

\item[{\rm 5.}]
$\g_k+u\leq \g_{k+1}$.

\item[{\rm 6.}]
$\g_k+\g_j+\g_k\g_j\leq \g_{k+j}$. \eproof
\end{enumerate}
\end{proposition}

From now on, whenever we write an expression containing $\theta_k$ we mean
that the same expression is true for some $\theta_k$, with
$|\theta_k| \le \g_k$.

When computing an arithmetic expression $q$ with a round-off
algorithm, errors will accumulate and we will obtain another
quantity which we will denote by $\fl(q)$. We write
$\Error(q)=|q-\fl(q)|$.

An example of round-off analysis which will be useful in what
follows is given in the next proposition, the proof of which can be
found in Section~3.1 of~\cite{Higham96}.

\begin{proposition}\label{scalar}
There is a round-off algorithm which, with input $x,y\in\R^n$,
computes the dot product of $x$ and $y$. The computed value
$\fl(\langle x,y\rangle)$ satisfies
$$
   \fl(\langle x,y\rangle)=\langle x,y\rangle+
   \t_{\lceil\log_2 n\rceil+1}\langle |x|,|y|\rangle,
$$
where $|x|=(|x_1|,\ldots,|x_n|)$. In particular, if $x=y$, the
algorithm computes $\fl(\|x\|^2)$ satisfying
\begin{equation}\tag*{\qed}
   \fl(\|x\|^2)=\|x\|^2(1+\t_{\lceil\log_2 n\rceil+1}).
\end{equation}
\end{proposition}

We will also have to deal with square roots and arccosinus. The
following result will help us to do so.

\begin{lemma}\label{sqrtLemma}
\begin{description}
\item[(i)]
Let $\t\in\R$ such that $|\t|\leq 1/2$. Then, $\sqrt{1-\t}=1-\t'$
with $|\t'|\leq |\t|$.
\item[(ii)]
Let $0<a\leq 1$ and $\e\in\R$ such that $0<a+\e <1$. Then,
$\arccos(a+\e)=\arccos(a)+\upsilon\frac{1}{\sqrt{1-(a+\e)^2}}$ with
$|\upsilon|\leq |\e|$.
\end{description}
\end{lemma}

\proof Assume $\t>0$ (if $\t<0$ it is done similarly). By the
intermediate value theorem we have that
$1-\sqrt{1-\t}=\t(\sqrt{\xi})'$ with $\xi\in(1-\t,1)$. But
$$
   (\sqrt{\xi})'
  =\frac{1}{2\sqrt{\xi}}\leq \frac{1}{\sqrt2},
$$
the last since $\xi\geq 1/2$. This proves~(i).

Part~(ii) is shown similarly.  Again, assume for simplicity that
$\e>0$. Then, for some $\xi\in(a,a+\e)$,
\begin{equation}\tag*{\qed}
 \arccos(a+\e)-\arccos(a)=\e\arccos'({\xi})
  =\e \frac{1}{\sqrt{1-\xi^2}}
  = \frac{\upsilon}{\sqrt{1-(a+\e)^2}}.
\end{equation}
\medskip

We assume that, besides the four basic arithmetic operations, we are
allowed to compute square roots and arccosinus with finite
precision. That is, if $\op$ denotes any of these two operators, we
compute $\tilde{\op}$ such that
$$
    \tilde{\op}(x)=\op(x)(1+\d),
    \quad |\d|\leq u.
$$
From Lemma~\ref{sqrtLemma}(i) it follows that, for all $a>0$,
$$
  \widetilde{\sqrt{a(1+\t_k)}}=\sqrt{a}(1+\t_{k+1}).
$$

\begin{remark}\label{precision}
Our choice of the precision $u$ in Theorem~\ref{th:main}(4)
guarantees that $ku<1/2$ holds whenever we encounter $\t_k$ in what
follows, and consequently, $|\t_k| \leq \gamma_k\leq 2ku$. This
implies that in all what follows   we have $\g_g=\Oh(ug)$ for all
the expressions $g$ we will encounter.
\end{remark}

According to the previous remark we will introduce a further
notation that will considerably simplify our exposition. For all
expression $g$, we will write
$$
    \ll g\rr:= \Oh(ug).
$$
This notation will avoid we burden ourselves with the consideration
of multiplicative constants.

\subsection{The finite precision algorithm}

Our finite precision algorithm is a variation of Algorithm {\tt
Count\_Roots\_1} in Section~\ref{sec:proof1}. Given $x\in S^n$ we
define below  $\fl({A'}(f))$ and $ \fl(\bar B'_f(x))$, which are
convenient floating versions of the sets ${A'}(f)=\left\{x\in
S^n\mid
   \bar\alpha(f,x)<\frac12\ab\right\}$ and $\bar B'_f(y)=\{z\in S^n \mid d(x,y)\le \frac{3}{2}\sigma
\bar{\beta}(f,x)\}$  respectively.

Given $f\in \Hd$ and $x\in S^n$, we let $M\in\R^{n\times n}$ be a
matrix representing
$$
  \left[\begin{matrix} \frac{1}{\sqrt{d_1}} \\ &
  \frac{1}{\sqrt{d_2}} \\
   & & \ddots \\ & & & \frac{1}{\sqrt{d_n}}
  \end{matrix} \right]Df(x)_{|T_x S^n}.
$$
and we set $\sigma_{\min}(M)=\|M^{-1}\|^{-1}$. Therefore
\begin{eqnarray*}
\mu_{\rm norm}(f,x)&=&\|f\|\sqrt n\, \|M^{-1}\|\ = \ \|f\|\sqrt n
\,\sigma_{\min}(M)^{-1},\\[2mm]
\bar\beta(f,x)&= & \mu_{\rm norm}(f,x)\frac{\|f(x)\|_\infty}{\|f\|}
\ =\ \sqrt n
\,\sigma_{\min}(M)^{-1}\|f(x)\|_\infty,\\
\bar\alpha(f,x)&= &\bar\beta(f,x) \mu_{\rm norm}(f,x)
\frac{\bD^{3/2}}{2}  = \|f\|\, n\,\sigma_{\min}(M)^{-2}
\|f(x)\|_{\infty} \frac{\bD^{3/2}}{2}.
\end{eqnarray*}
This implies that $$\begin{array}{ccccc} y\in \bar B'_f(x) &\iff&
d(x,y)\le \frac32\sigma\bar{\beta}(f,x) &\iff & \sigma_{\min}(M)
d(x,y)\le \frac32 \sigma \sqrt n
\|f(x)\|_\infty,\\[2mm]
x\in A'(f) &\iff& \bar\alpha(f,x)<\frac\ab 2 & \iff & \|f\|
n\|f(x)\|_\infty\bD^{3/2}<\ab\sigma_{\min}(M)^2.
\end{array}$$
These statements are equivalent under infinite precision, but the 
expressions at the right-hand side are
more convenient to handle when working with finite precision. This
motivates our definitions of
\begin{eqnarray*}
\fl(\bar B'_f(x))&:=&\left\{ y\in S^n \mid
\fl(\sigma_{\min}(M)d(x,y))\leq \fl(\frac32 \sigma \sqrt n
\|f(x)\|_\infty )\right\}\\
  \fl({A'}(f))&:=& \left\{x\in S^n\mid
   \fl(\|f\|\,n\,\|f(x)\|_\infty\bD^{3/2})<\fl(\ab\sigma_{\min}(M)^2)\right\}
   \end{eqnarray*}

We also define accordingly the graph $\fl(G'_\eta)$ whose  
vertices are the points in $G_\eta \cap \fl(A'(f))$, and with two
vertices $x,y$  joined by an edge if and only if $\fl(\bar
B'_f(x))\cap \fl(\bar B'_f(x))\ne \emptyset$. Its connected components
are denoted by $\fl(U)$.

\bigskip Our algorithm is the following:

{\small \algo
\>\> {\tt Count\_Roots\_2}$(f)$\\
\>\> let $\eta:=\frac{2\sqrt{2}}{\pi\sqrt{n+1}}$\\
(1)  \>\> let $\fl(U_1),\ldots,\fl(U_r)$ be the connected components
of
      $\fl(\bar{G}_\eta)$\\
\>\> if \\
\>\>\> (i) for $1\leq i<j\leq r$ \\
\>\>\>\> for all $x_i\in \fl(U_i)$ and all $x_j\in \fl(U_j)$,
  $\fl(d(x_i,x_j))>\fl(\frac{3}{2}\pi \eta\sqrt{n+1})$ \\  %era un 2 antes%
\>\> and \\
\>\>\> (ii) for all $x\in\GG_\eta\setminus \fl(A'(f))$,
    $\fl(\|f(x)\|_\infty)> \fl(\frac{\sqrt 2}{2}\pi \eta\sqrt{(n+1)\bD}\|f\|)$\\
\>\> then HALT and return $r/2$\\
\>\> else $\eta:=\eta/2$ \\
\>\>\> go to (1) \falgo }

In the rest of the section we will see that, when the precision $u$
satisfies  $u\leq\frac{1}{\Oh\left(\bD^2n^{5/2}\kappa(f)^3(\log
S+n^{3/2} \bD^2\kappa(f)^2)\right)}$, this algorithm is correct and
halts as soon as $\eta\le  \frac{\ab}{4\bD^2(n+1)\kappa(f)^2}$.

\subsection{Bounding errors for elementary computations}
\label{sec:fpc}

The goal of this subsection is to exhibit bounds for the accumulated
error in the main computations of {\tt Count\_Roots\_2}. We
will rely on the basic notations and results described
in~\S\ref{subsec:bf}.

To simplify notation, and without loss of generality, in all what
follows we assume that $\|f\|=1$. We denote by $S(\Hd)$ the sphere
of such systems. Also, we do not discuss in what follows the
accumulated error in the computation of $\phi:C^n\to S^n$. This is a
minor detail which can be taken care of using
Lemma~\ref{sqrtLemma}(i).

\begin{proposition}\label{prop:error1}
Given $f\in S(\Hd)$ and $x\in S^n$, we can compute $\|f(x)\|_\infty$
with finite precision $u$ such that
$$
  \Error(\|f(x)\|_\infty)= \ll\bD+\log S\rr
$$
where $S$ is a bound on the number of coefficients of each $f_i$.
\end{proposition}

\proof Let $f=(f_1,\ldots,f_n)$. For $i\leq n$ write $f_i=\sum c_J
X^J$ and let $S$ be the number of coefficients of $f_i$. To compute
$f(x)$ one computes each monomial $c_J x^J$ with
$\fl(c_Jx^J)=c_Jx^J(1+\t_D)$. Then, one computes $f_i(x)$ to get
\begin{eqnarray*}
 \fl(f_i(x)) &=& \fl(\sum \fl(c_Jx^J))\\
   &=& \fl(\sum c_Jx^J (1+\theta^{(J)}_\bD))\\
   &=& \sum c_Jx^J (1+\theta^{(J)}_\bD) +
       \theta_{\log S}\sum |c_Jx^J| (1+\theta^{(J)}_\bD)\\
   &=& f_i(x)  + \sum c_Jx^J\theta^{(J)}_\bD +
      \theta_{\log S}\sum |c_Jx^J| (1+\theta^{(J)}_\bD)
\end{eqnarray*}
where in the third line we reasoned as in the proof of
Proposition~\ref{scalar}. Therefore
\begin{eqnarray*}
 \Error(\|f(x)\|_\infty)&\le & \Big| \sum c_Jx^J\theta^{(J)}_\bD +
      \theta_{\log S}\sum |c_Jx^J| (1+\theta^{(J)}_\bD)\Big|\\
      &\le & \sum |c_J|\,\|x^J\| (\g_\bD+\g_{\log S} +
      \g_{\bD}\g_{\log S})\\
      &\le & \g_{\bD+\log S}
\end{eqnarray*}
where we used that for any $x\in S^n$, $| \sum |c_J| x^J|\le \|\sum
|c_J|x^J\| =\|f_i\| \leq \|f\|=1$ and
Proposition~\ref{propHigham2}~(6). The conclusion follows from
Remark~\ref{precision}. \eproof

\begin{proposition}\label{prop:g1}
Given $f\in S(\Hd)$ and $x\in S^n$, let $M\in\R^{n\times n}$ be a
matrix representing
$$
  \left[\begin{matrix} \frac{1}{\sqrt{d_1}} \\ &
  \frac{1}{\sqrt{d_2}} \\
   & & \ddots \\ & & & \frac{1}{\sqrt{d_n}}
  \end{matrix} \right]Df(x)_{|T_x S^n}
$$
in some orthonormal basis of $T_x S^n$.
Then $\|M\|\leq\sqrt{n}$. In addition,
we can compute such a matrix 
$M$ with finite precision $u$ such that
$$
    \|\Error(M)\|_F=\ll n (\log S + \bD+\log n)\rr.
$$
\end{proposition}

\proof 
{\bf Step 1:}
Let $y = \frac{x - \mathrm e_{n+1}}{\|x - \mathrm e_{n+1}\|}$. The
{\em Householder symmetry}
\[
    H_y = I_{n+1} - 2 y y^t
\]
swaps vectors $\mathrm e_{n+1}$ and $x$, and fixes $y^{\perp}$. The
first $n$ columns of $H_y$ are therefore an orthonormal basis of
$T_xS^n$, while the last column is $x$. Let $ H\in \R^{(n+1)\times
n}$ denote the submatrix obtained from the first $n$ columns of
$H_y$. With that notation, we set
\[
M =
  \left[\begin{matrix} \frac{1}{\sqrt{d_1}} \\ &
  \frac{1}{\sqrt{d_2}} \\
   & & \ddots \\ & & & \frac{1}{\sqrt{d_n}}
  \end{matrix} \right]Df(x)  H.
\]
\\
{\bf Step 2:}
We claim that $P_{i,x}: \mathcal H_{d_i}
\rightarrow \mathbb R^n$,
$f_i \mapsto \frac{1}{\sqrt{d_i}} Df_i(x)_{|T_xS^n}$ is an orthogonal
projection, in the sense that for any fixed $x$, the map
$(P_{i,x})_{| \ker (P_{i,x}) ^{\perp}}$ is
an isometry.

We  use an orthogonal invariance argument. The special orthogonal
group $SO(n+1)$ acts on $\mathcal H_{d_i}$ and on $\mathbb R^{n+1}$
isometrically as follows: to a given $Q \in SO(n+1)$, we associate
respectively the following isometries:
\[
  x   \mapsto   Q x \quad , \quad  f_i   \mapsto   f_i \circ Q^t  .
\]
We set $y = Qx$ and $g_i = f_i \circ Q^t$. Differentiating the
equality $g_i(Q x) = f_i(x)$, we obtain:
\[
  Dg_i(y) Q = Df_i(x) .
\]
When $x$ is fixed, we can set $Q$ conveniently so that 
$y=\mathrm e_{n+1}$. Therefore
\[
  Dg_i(\mathrm e_{n+1}) Q_{|T_xS^{n}} = Df_i (x)_{|T_xS^{n}}  .
\]
Since $Q(T_xS^{n})=T_{\mathrm e_{n+1}}S^{n}$ we obtain
\[
  Dg_i(\mathrm e_{n+1})_{|T_{\mathrm e_{n+1}}S^{n}} 
  = Df_i (x)_{|T_xS^{n}}  .
\]
This means that $P_{i,\mathrm e_{n+1}} (f_i \circ Q^t) = P_{i,x} (f_i)$.
Thus, in order to prove our claim, it is enough to show that
$P_{i,\mathrm e_{n+1}}$ is an orthogonal projection.

Since for $g=\sum_J g_JX^J$, $\frac{\partial g}{\partial X_j}
(\mathrm e_{n+1})=g_{(\mathrm e_j+(d-1)\mathrm e_{n+1})}$ 
and since $T_{e_{n+1}}S^n =\langle
\mathrm e_1,\dots,\mathrm e_n\rangle$, 
we have that for any $g_i\in \mathcal H_{d_i}$,
\[
 P_{i, \mathrm e_{n+1}} (g_i) = 
  \frac{1}{\sqrt {d_i}} \big(g_{i (\mathrm e_1
 + (d_i-1) \mathrm e_{n+1})} ,   \dots , g_{i  (\mathrm e_n + (d_i-1)
 \mathrm e_{n+1}) } \big) \ .
\]
Hence, for any $g_i\in   \ker (P_{i,\mathrm e_{n+1}}) ^{\perp}$, i.e. such
that $g_{iJ}=0 $ for all $J\ne \mathrm e_j + (d_i-1) \mathrm
e_{n+1}$, $1\le j\le n$, we have
$$
  \|g_i\|^2 = \sum_J \frac{g_{iJ}^2}{{d_i \choose J}}=  
  \|P_{i, \mathrm e_{n+1}} (g_i)\|_2^2 .
$$
We conclude that  $P_{i,x}$ is an orthogonal projection.

\smallskip \noindent
{\bf Step 3:} From the previous step, for any $f_i\in \mathcal
H_{d_i}$, using the orthogonal decomposition $f_i= f_i^\circ+
f_i^\perp$ with $f_i^\circ \in \ker P_{i,x}$ and $f_i^\perp \in \ker
P_{i,x}^\perp$, we have
$$
   \|P_{i,x}(f_i)\|_2^2 = \|P_{i,x}(f_i^\perp)\|_2^2 
  =\|f_i^\perp\|^2\le  \|f_i\|^2.
$$ 
It is now immediate from Step 1 and from the definition of 
$\|f\|=\max_i\|f_i\|$ that
the Frobenius norm $\|M\|_F$ of the matrix $M$ satisfies
\[
  \|M\|_F^2 = \sum_{i=1}^n \|P_{i,x}(f_i)\|_2^2 \le  
  \sum_{i=1}^n \|f_i\|^2  \le {n} \|f\|^2 = {n}
\]
and hence its spectral norm $\|M\|$ satisfies $\|M\|\le \|M\|_F\le
\sqrt{n}$. This bound is independent of the choice of the basis
for the space $T_xS^n$.

\smallskip \noindent
 {\bf Step 4:} 
We next present the algorithm to compute $M$, given $f$ and $x$. This is 
a non-optimal algorithm, and can be significantly improved if more is
known on the structure of the polynomial system $f$.

We can compute each entry $m_{ij}$ of the matrix $M$ as the scalar product
of $\frac{1}{\sqrt d_i} D f_i (x) $ and the $j$th column
$H_j:=(h_{kj})_{1\le k\le n+1}$ of $H$.

Proceeding as in the proof of Proposition~\ref{prop:error1}, we can
compute $ \frac{1}{\sqrt d_i}\,\frac{\partial f_i}{\partial X_k}
(x)$ with
$$  
  \Error\left(\frac{1}{\sqrt d_i}\,\frac{\partial f_i}{\partial X_k}
     (x)\right)= \ll\bD+\log S\rr.
$$
On the other hand, the vector $y = \frac{x - \mathrm e_{n+1}}{\|x -
\mathrm e_{n+1}\|}$ can be computed using  
$2n+4$ operations,
and clearly $\Error(y_j) = \ll \log (n) \rr$ for all $j$. Hence, 
for all coefficients $h_{kj}$ of $H$, 
$$
  \Error (h_{kj})= \ll \log (n) \rr.
$$
Applying Proposition~\ref{scalar} we conclude
\begin{eqnarray*}\Error (m_{ij})&=& \ll\bD+\log S + \log n \rr
\left\|\frac{1}{\sqrt {d_i}}Df_i(x)\big) \right\| \,\|H_j\|\\
&=& \ll \bD+\log S + \log n\rr.
\end{eqnarray*}
The second equality holds because
$\|H_j\|= 1$ since $ H$ is unitary, and
because, as in the proof of Step 2,
$$
  \left\|\frac{1}{\sqrt {d_i}}D f_i(x)\right\|^2= \left
 \|\frac{1}{\sqrt {d_i}}D g_i(\mathrm e_{n+1})\right\|^2=\frac{1}{ {d_i}}
 \|(g_{i (\mathrm e_1 + (d_i-1) \mathrm e_{n+1})} ,   \dots , g_{i
 (d_i\mathrm e_{n+1}) })\|^2  \le \|g_i\|^2\le 1.
$$
This implies
\begin{equation}\tag*{\qed}
 \| \Error (M)\|_F \le \ll n(\bD+\log S+\log n) \rr.
\end{equation}

\begin{lemma}\label{lem:inverse}
Let $x\in S^n$ and $M$ be as in
Proposition~\ref{prop:g1}. We can compute
$\sigma_{\min}(M)=\|M^{-1}\|^{-1}$ satisfying
$$
    \Error(\sigma_{\min}(M))= \ll n (\log S + \bD + n^{3/2})\rr.
$$
\end{lemma}
% Abajo ya modifiqué, sacando el factor de n que aparecia en prop:g1
\proof
Let $E'=M-\fl(M)$. By Proposition~\ref{prop:g1},
$$
  \|E'\|\leq \|E'\|_F\leq \ll n (\log S + \bD +\log n)\rr.
$$
Let $\MM=\fl(M)$. We compute
$\sigma_{\min}(\MM)=\|M^{-1}\|^{-1}$ using a backward stable algorithm
(e.g., QR factorization).
Then the computed $\fl(\sigma_{\min}(\MM))$ is the exact
$\sigma_{\min}(\MM+E'')$ for a matrix $E''$ with
\[
  \|E''\| \leq cn^2 u \|\MM\|
\]
for some universal constant $c$ (see, e.g., \cite{GoLoan,Higham96}).
Thus,
$$
  \fl(\sigma_{\min}(M))=\fl(\sigma_{\min}(\MM))=
  \sigma_{\min}(\MM+E'')=\sigma_{\min}(M+E'+E'').
$$
Write $E=E'+E''$. Then, using $\|M\| \leq \sqrt{n}$,
\begin{eqnarray*}
  \|E\| &\leq& \|E'\|+\|E''\|
  \leq \|E'\| + cn^2 u \|\MM\|
  \leq \|E'\| + cn^2 u (\|M\|+\|E'\|)\\
  &=& \ll n (\log S + \bD +\log n)\rr
   + cn^2 u (\sqrt{n}+\ll n (\log S + \bD +\log n)\rr)\\
   &=& \ll n (\log S + \bD +\log n)\rr 
  + cn^2 u(\sqrt{n} +c'u n(\log S+\bD+n^{3/2})) \\
  &=& \ll n (\log S + \bD + n^{3/2})\rr
\end{eqnarray*}
since the hypothesis on $u$ implies $c' u n(\log S+\bD+n^{3/2})$ is
bounded by a constant term.

Therefore, $\fl(\sigma_{\min}(M)) = \sigma_{\min}(M+E)$ which
implies by \cite[Corollary~8.3.2]{GoLoan}:
\begin{equation}\tag*{\qed}
  \Error(\sigma_{\min}(M))\leq \|E\|
  <\ll n (\log S + \bD + n^{3/2})\rr.
\end{equation}

\begin{proposition}\label{prop:alphabar}
Let $f\in S(\Hd))$. Assume $u\leq\frac{K}{\kappa(f)^2n^2\bD\log S}$
for a small enough constant and let $x\in S^n$. Then
\begin{description}
\item[(i)]
If $x\notin \fl(A'(f))$ then
%If $\fl(\bar\alpha(f,x))\geq\frac12\ab$ then
$\bar\alpha(f,x)\geq\frac{1}{3}\ab$.
\item[(ii)]
If $x\in \fl(A'(f))$ then $\bar\alpha(f,x)<\ab$.
\end{description}
\end{proposition}

\proof
From Proposition~\ref{prop:error1} 
\begin{eqnarray*}
   \fl(n\|f(x)\|_\infty\bD^{3/2}) &=&
   (\|f(x)\|_\infty+\ll\bD+\log S\rr)
   (n\bD^{3/2})(1+\theta_4) \\
   &\leq& n\bD^{3/2}\|f(x)\|_\infty +
     \ll n\bD^{3/2}(\bD+\log S)\rr
\end{eqnarray*}.
Also, from Lemma~\ref{lem:inverse}, 
using that $\sigma_{\min}(M)\leq\sqrt{n}$,
\begin{eqnarray*}
 \fl(\ab\sigma_{\min}(M)^2)
 &=& \ab\left(\sigma_{\min}(M)
  +\ll n(\log S + \bD + n^{3/2})\rr\right)^2 (1+\theta_2)\\
  &\ge& \ab \sigma_{\min}(M)^2 -2\ab \sigma_{\min}(M)
\ll n (\log S + \bD + n^{3/2})\rr  \\
&\ge & \ab\sigma_{\min}(M)^2 - \ll n^{3/2} (\log S + \bD +
n^{3/2})\rr.
\end{eqnarray*}
Therefore,
\begin{eqnarray*}
  n\|f(x)\|_\infty\bD^{3/2}+\ll n\bD^{3/2}(\bD+\log S)\rr
 &\geq&
   \fl(n\|f(x)\|_\infty\bD^{3/2}) \geq
   \fl(\ab\sigma_{\min}^2)\\
 &\geq& \ab\sigma_{\min}^2 - \ll n^{3/2}(\log S + \bD + n^{3/2})\rr
\end{eqnarray*}
or yet,
%using that $\fl(\bar\alpha(f,x))\geq \frac12\ab$,
\begin{eqnarray*}
  n\|f(x)\|_\infty\bD^{3/2}-\ab\sigma_{\min}^2 &\geq&
  -(\ll n\bD^{3/2}(\bD+\log S)\rr +\ll n^{3/2}
   (\log S + \bD + n^{3/2})\rr)\\
  &\geq& - \ll n^{3}\bD^{5/2}\log S\rr.
\end{eqnarray*}
\smallskip

\noindent
\fbox{{\bf Case I.}
$\min\left\{\mu_{\rm norm}(f,x),\frac{1}{\|f(x)\|_\infty}\right\}
=\frac{1}{\|f(x)\|_\infty}$}\medskip

\noindent
In this case $\kappa(f)\geq \frac{1}{\|f(x)\|_\infty}$ and,
therefore, using the hypothesis on $u$ and the inequality $\kappa(f)\geq 1$,
\begin{eqnarray*}
   \ll n^{3}\bD^{5/2}\log S\rr
   &=& u \Oh(n^{3}\bD^{5/2}\log S)
   \leq K \frac{\Oh(n^{3}\bD^{5/2}\log S)}{\kappa(f)n^{2}\bD\log S}\\
   &\leq& K\Oh(1) n\|f(x)\|_\infty\bD^{3/2}
      \leq\frac{n\|f(x)\|_\infty\bD^{3/2}}{2}
\end{eqnarray*}
the last by choosing $K$ small enough. Hence,
$n\|f(x)\|_\infty\bD^{3/2}-\ab\sigma_{\min}^2
 \geq -\left(\frac{n\|f(x)\|_\infty\bD^{3/2}}{2}\right)$,
which implies $\frac{3}{2}n\|f(x)\|_\infty\bD^{3/2} \geq
\ab\sigma_{\min}(M)^2$, i.e., $\bar\alpha(f,x)\geq\frac{\ab}{3}$.
\medskip

\noindent
\fbox{{\bf Case II.}
$\min\left\{\mu_{\rm norm}(f,x),\frac{1}{\|f(x)\|_\infty}\right\}
=\mu_{\rm norm}(f,x)$}\medskip

\noindent
In this case
$\kappa(f)\geq \mu_{\rm norm}(f,x)=\frac{\sqrt{n}}
{\sigma_{\min}(M)}$.
By the hypothesis on $u$,
\begin{eqnarray*}
   \ll n^{3}\bD^{5/2}\log S\rr
   &=& u \Oh(n^{3}\bD^{5/2}\log S)
   \leq K \frac{\Oh(n^{3}\bD^{5/2}\log S)}{\kappa(f)^2n^{2}\bD\log S}\\
   &\leq& K\Oh(1) \sigma_{\min}(M)^2\bD^{3/2}
      \leq \frac{\ab\sigma_{\min}(M)^2}{3}
\end{eqnarray*}
the last by choosing $K$ small enough. This implies
$n\|f(x)\|_\infty\bD^{3/2}-\ab\sigma_{\min}(M)^2 \geq
-\frac{\ab\sigma_{\min}(M)^2}{3}$ or, equivalently,
$\bar\alpha(f,x)\geq\frac{\ab}{3}$.
\medskip

This shows part~(i). For part~(ii), one shows as above that
$$
  n\|f(x)\|_\infty\bD^{3/2}-\ab\sigma_{\min}^2 \leq
  \ll n^{3}\bD^{5/2}\log S\rr.
$$
Then, one proceeds as well by considering the two cases
$\min\left\{\mu_{\rm norm}(f,x),\frac{1}{\|f(x)\|_\infty}\right\}
=\frac{1}{\|f(x)\|_\infty}$ and
$\min\left\{\mu_{\rm norm}(f,x),\frac{1}{\|f(x)\|_\infty}\right\}
=\mu_{\rm norm}(f,x)$.
\eproof

\begin{lemma}\label{lem:psi}
Let $y_1,y_2\in \UU_\eta$ and let $x_i=\phi(y_i)$, $i=1,2$. Then
$d(x_1,x_2)\geq \frac{\eta}{2\sqrt{n+1}}$.
\end{lemma}

\proof
The distance $d(x_1,x_2)$ is minimized at
$y_1=(1,\ldots,1,1)$ and $y_2=(1,\ldots,1,1-\eta)$. Let
$N=n+1$. Then
\begin{eqnarray*}
  \cos(d(x_1,x_2))^2&=& \frac{\langle y_1,y_2\rangle^2}
        {\|y_1\|^2\|y_2\|^2}\\
   &=& \frac{(N-\eta)^2}{N(N-2\eta+\eta^2)}\\
   &=& 1-\frac{(N-1)\eta^2}{N^2-2N\eta+N\eta^2}\\
   &\leq& 1-\eta^2\frac{N-1}{N^2}.
\end{eqnarray*}
Hence
$$
  d(x_1,x_2)\geq \arccos\left(\sqrt{1-\eta^2\frac{N-1}{N^2}}
   \right) =\arcsin \left(\frac{\eta}{N}\sqrt{N-1}\right)
   \geq \frac{\eta}{2\sqrt N}.
$$
\eproof

\begin{lemma}\label{lem:dist}
Let $u<\frac{K\eta^2}{n\log n}$ for a small
enough constant $K$.
For $x_1,x_2\in \GG_\eta$
we can compute $d(x_1,x_2)$ such that
$$
   \Error(d(x_1,x_2))\leq \bll\frac{\sqrt{n}\log n}{\eta}\brr.
$$
\end{lemma}

\proof Let $y_i=\phi^{-1}(x_i)$, $i=1,2$, and $a=\cos(d(x_1,x_2))$,
i.e.,
$$
   a=\frac{\langle y_1,y_2\rangle}{\|y_1\|\|y_2\|}.
$$
We have, using Proposition~\ref{scalar},
$$
 \fl(\langle y_1,y_2\rangle)=\langle y_1,y_2\rangle
 +\theta_{\log n}\|y_1\|\|y_2\|
$$
and $\fl(\|y_1\|\|y_2\|)=\|y_1\|\|y_2\|(1+\theta_{\log n})$.
Using now Propositions~\ref{prop:f1},~\ref{propHigham},
and~\ref{propHigham2},
it follows that $\fl(a)=a+\e$ with $\e=\ll\log n\rr$.

By choosing $K$ sufficiently small,
$\e\leq\frac{\eta^2n}{12(n+1)^2}$. Also, from the proof of
Lemma~\ref{lem:psi},
$$
  a=\cos(d(x_1,x_2))\leq
   \sqrt{1-\frac{\eta^2n}{(n+1)^2}}
$$
and hence, using that $\sqrt{z}+y \leq \sqrt{z+3y}$ whenever
$0<z,y\leq 1$, we obtain
$$
  a+\e \leq
  \sqrt{1-\frac{\eta^2n}{(n+1)^2}} +\frac{\eta^2n}{12(n+1)^2}
  \leq \sqrt{1-\frac{3\eta^2n}{4(n+1)^2}}
  \leq \sqrt{1-\frac{\eta^2}{3(n+1)}}.
$$
Using Lemma~\ref{sqrtLemma}(ii) it follows that,
\begin{eqnarray*}
  \arccos(a+\e)
  &=& \arccos(a)+ \e\left|\frac{1}
  {\sqrt{1-(a+\e)^2}}\right|\\
  &=& \arccos(a)+ \ll\log n\rr\left|\frac{\sqrt{3(n+1)}}
  {\eta}\right|.
\end{eqnarray*}
Therefore,
\begin{equation}\tag*{\qed}
  \Error(d(x_1,x_2))\leq
   \bll\frac{\sqrt{n}\log n}{\eta}\brr.
\end{equation}
\medskip

\begin{lemma}\label{lem:beta} Let $f\in
S(\Hd)$. Assume that $\eta\geq \frac{\ab}{8\bD^2(n+1)\kappa(f)^2}$
and $u\leq\frac{K}{\bD^2n^{5/2}\kappa(f)^3(\log
S+n^{3/2}\bD^2\kappa(f)^2)}$ with $K$ small enough, and  let
$x,y\in\GG_\eta$. Then
\begin{description}
\item[(i)]
If $y\in \fl(\bar B'_f(x))$ then $d(x,y)\le
2\sigma\bar{\beta}(f,x)$.
\item[(ii)]
If $y\notin \fl(\bar B'_f(x))$ then $d(x,y)>\sigma\bar{\beta}(f,x)$.
\end{description}
\end{lemma}

\proof
By Lemmas~\ref{lem:inverse} and~\ref{lem:dist} (and using
$\sigma_{\min}(M)\leq \sqrt{n}$ and the bound
$d(x,y)\leq \frac{\pi}{2}\eta\sqrt{n+1}$ which follows from~\eqref{eq:b1}),
\begin{eqnarray*}
   \Error(\sigma_{\min}(M)d(x,y))&=&
   \Oh\bigl(d(x,y)\Error(\sigma_{\min}(M))
      +\sigma_{\min}(M)\Error(d(x,y))\bigr)\\
  &=&\eta\frac{\pi}{2}\sqrt{n+1}\ll n(\log S+\bD+n^{3/2})\rr
   +\sqrt{n}\bll\frac{\sqrt{n}\log n}{\eta}\brr\\
 &=& \eta\ll n^{3/2}(\log S+\bD+n^{3/2})\rr
   +\bll\frac{n\log n}{\eta}\brr\\
 &\leq& \ll n^{3/2}\log S+n^3\bD^2\kappa(f)^2\rr
\end{eqnarray*}
the last by the bounds on $\eta$.
Also, using Proposition~\ref{prop:error1},
$$
   \Error\left(\frac32\sigma\sqrt{n}\|f(x)\|_\infty\right)
   \leq \ll\sqrt{n}(\bD+\log S)\rr.
$$
Therefore, for part~(i),
\begin{align*}
  \sigma_{\min}(M)&d(x,y)-\frac32\sigma\sqrt{n}\|f(x)\|_\infty\\
  \le \;& \fl(\sigma_{\min}(M)d(x,y))-
  \fl\left(\frac32\sigma\sqrt{n}\|f(x)\|_\infty\right)
 +\ll n^{3/2}\log S+n^3\bD^2\kappa(f)^2\rr
   +\ll\sqrt{n}(\bD+\log S)\rr\\
 \leq\;&\ll n^{3/2}\log S+n^3\bD^2\kappa(f)^2\rr
   +\ll\sqrt{n}(\bD+\log S)\rr\\
 =\;& \ll n^{3/2}\log S+n^3\bD^2\kappa(f)^2\rr.
\end{align*}
\smallskip

\noindent
\fbox{{\bf Case I.}
$\min\left\{\mu_{\rm norm}(f,x),\frac{1}{\|f(x)\|_\infty}\right\}
=\frac{1}{\|f(x)\|_\infty}$}\medskip

\noindent In this case $\kappa(f)\geq \frac{1}{\|f(x)\|_\infty}$
and, therefore, by the hypothesis on $u$,
\begin{eqnarray*}
  \ll n^{3/2}\log S+n^3\bD^2\kappa(f)^2\rr
  &=&\Oh(n^{3/2}\log S+n^3\bD^2\kappa(f)^2)
     \frac{K}{\kappa(f)n(\log S+n^{3/2}\bD^2\kappa(f)^2)}\\
 &\leq&\frac{\sigma\sqrt{n}}{2\kappa(f)}
 \leq\frac{\sigma\sqrt{n}\|f(x)\|_\infty}{2}
\end{eqnarray*}
the last line by taking $K$ small enough. This implies that
$\sigma_{\min}(M)d(x,y)\le  2\sigma\sqrt{n}\|f(x)\|_\infty$, i.e.,
that $d(x,y)\le  2\sigma\bar{\beta}(f,x)$.
\medskip

\noindent
\fbox{{\bf Case II.}
$\min\left\{\mu_{\rm norm}(f,x),\frac{1}{\|f(x)\|_\infty}\right\}
=\mu_{\rm norm}(f,x)$}\medskip

\noindent
In this case
$\kappa(f)\geq \mu_{\rm norm}(f,x)=\frac{\sqrt{n}}
{\sigma_{\min}(M)}$.
By the hypothesis on $u$
\begin{eqnarray*}
  \ll n^{3/2}\log S+n^3\bD^2\kappa(f)^2\rr
  &=&\Oh(n^{3/2}\log S+n^3\bD^2\kappa(f)^2)
     \frac{K}{\bD^2n^{5/2}\kappa(f)^3(\log S+n^{3/2}\bD^2\kappa(f)^2)}\\
 &\leq&\frac{\sqrt{n}\ab}{48\bD^2(n+1)^{3/2}\kappa(f)^3}\\
 &\leq&\frac{\sqrt{n}\eta}{8\sqrt{n+1}\kappa(f)}
 \leq\frac{\sqrt{n}d(x,y)}{4\kappa(f)}
 \leq\frac{\sigma_{\min}(M)d(x,y)}{4}
\end{eqnarray*}
 by taking $K$ small enough and Lemma~\ref{lem:psi}.
This implies that $\frac34\sigma_{\min}(M)d(x,y)\leq
\frac32\sigma\sqrt{n}\|f(x)\|_\infty$, i.e., that $d(x,y)\leq
2\sigma\bar{\beta}(f,x)$.
\medskip

This shows part~(i). Part~(ii) is shown in a similar way.
\eproof

\begin{lemma}\label{lem:e1}
Let $u\leq \frac{K\eta^2}{\log n}$ with $K$ small enough
and $x_1,x_2\in \GG_\eta$.
\begin{description}
\item[(i)]
If $\fl(d(x_1,x_2))\leq \fl(\frac32\pi\eta\sqrt{n+1})$ then
$d(x_1,x_2)\leq 2\pi\eta\sqrt{n+1}$.
\item[(ii)]
If $\fl(d(x_1,x_2))>\fl(\frac32\pi\eta\sqrt{n+1})$ then
$d(x_1,x_2)>\pi\eta\sqrt{n+1}$.
\end{description}
\end{lemma}

\proof
By Lemma~\ref{lem:dist} and the hypothesis on $u$, we obtain
$$
  \Error(d(x_1,x_2))=\bll\frac{\sqrt{n}\log n}{\eta}\brr
    \leq\Oh\left(\frac{\sqrt{n}\log n}{\eta}\right)
    \frac{K\eta^2}{\log n}
    \leq \frac{\pi}{2}\eta\sqrt{n+1},
$$
the last by taking $K$ small enough. Also,
$\Error(\frac32\pi\eta\sqrt{n+1})\leq
\frac32\pi\eta\sqrt{n+1}\;\gamma_3$. The statement easily follows
from these two bounds. \eproof

\begin{lemma}\label{lem:e2}
Let $u\leq\frac{K\eta\sqrt{n\bD}}{\bD+\log S+\eta\sqrt{n\bD}}$ with
$K$ small enough,
$f\in S(\Hd)$ and $x\in S^n$.
\begin{description}
\item[(i)]
If $\fl(\|f(x)\|_\infty) \leq \fl(\frac{\sqrt
2}{2}\pi\eta\sqrt{(n+1)\bD})$ then $\|f(x)\|_\infty \leq
\pi\eta\sqrt{(n+1)\bD}$.
\item[(ii)]
If $\fl(\|f(x)\|_\infty) >\fl(\frac{\sqrt
2}{2}\pi\eta\sqrt{(n+1)\bD})$ then $\|f(x)\|_\infty
>\frac{\pi}{2}\eta\sqrt{(n+1)\bD}$.
\end{description}
\end{lemma}

\proof
For part~(i), from Proposition~\ref{prop:error1},
$$
   \|f(x)\|_\infty
   \leq \fl(\|f(x)\|_\infty)+\ll\bD+\log S\rr.
$$
Also,
$$
  \frac{\sqrt
2}{2}\pi\eta\sqrt{(n+1)\bD}
   \geq \fl(\frac{\sqrt
2}{2}\pi\eta\sqrt{(n+1)\bD}) - \ll\eta\sqrt{(n+1)\bD}\rr.
$$
Therefore,
\begin{eqnarray*}
  \|f(x)\|_\infty- \frac{\sqrt
2}{2}\pi\eta\sqrt{(n+1)\bD} &\leq&
  \fl(\|f(x)\|_\infty) - \fl( \frac{\sqrt
2}{2}\pi\eta\sqrt{(n+1)\bD})
  + \ll\bD+\log S+\eta\sqrt{(n+1)\bD}\rr \\
  &\leq& \ll\bD+\log S+\eta\sqrt{(n+1)\bD}\rr\\
  &=& \Oh\bigl(\bD+\log S+\eta\sqrt{(n+1)\bD}\bigr)
     \frac{K\eta\sqrt{n\bD}}{\bD+\log S+\eta\sqrt{n\bD}}\\
  &\leq & (1-\frac{\sqrt 2}{2})\eta\sqrt{(n+1)\bD},
\end{eqnarray*}
the last by taking $K$ sufficiently small. It follows that
$\|f(x)\|_\infty \leq \pi \eta\sqrt{(n+1)\bD}$ and hence, part~(i)
of the statement.

Part~(ii) is proved similarly.
\eproof

\subsection{Proof of Theorem~\ref{th:main}(4): Correctness}\label{sec:corr}

We will show that, if
$u\leq\frac{1}{\Oh\left(\bD^2n^{5/2}\kappa(f)^3(\log S+n^{3/2}
\bD^2\kappa(f)^2)\right)}$, and the algorithm halts with $\eta \ge
\frac{\ab}{8\bD^2(n+1)\kappa(f)^2}$, then the value $r/2$ returned
by the algorithm is $\#_{\R}(f)$. This is a consequence of the
floating following versions of Lemmas~\ref{lem:D(x)} and
\ref{lem:!}.

\begin{lemma} \label{lem:fl51}
Let  $f\in S(\Hd)$,  $\eta \ge \frac{\ab}{8\bD^2(n+1)\kappa(f)^2}$
and $u\leq\frac{1}{\Oh\left(\bD^2n^{5/2}\kappa(f)^3(\log S+n^{3/2}
\bD^2\kappa(f)^2)\right)}$.

\begin{description}
\item[(i)]
For each $x\in \fl(A'(f))$ there exists  $\zeta_x\in Z(f)$ such that
$\zeta_x \in \bar B_f(x)$. Moreover for each point $z\in \fl(\bar
B'_f(x))$, the Newton sequence starting at $z$ converges to
$\zeta_x$.

\item[(ii)] Let $x,y \in \fl(A'(f))$. Then
$\zeta_x=\zeta_y \iff  \fl(\bar B'_f(x))\cap \fl(\bar
B'_f(y))\neq\emptyset$.
\end{description}
\end{lemma}

\proof  (i) Applying Proposition~\ref{prop:alphabar}(ii), $x\in
\fl(A'(f))$ implies that $\bar\alpha(f,x)<\ab$. Therefore, by
Theorem~\ref{th:alpha1}, there exists $\zeta_x\in Z(f) $ such that
$\zeta_x\in \bar B_f(x)$. Moreover, if $z\in \fl(\bar B'_f(x))$, by
Lemma~\ref{lem:beta}(i), $d(x,z)\le 2\sigma\bar\beta(f,x)$ and the
Newton sequence starting at $z$ converges to $\zeta_x$. \\
(ii) If $\zeta_x =  \zeta_y $, then $\bar B_f(x)\cap \bar B_f(y) \ne
\emptyset $ which implies by Lemma~\ref{lem:beta}(ii) that there
exists $z\in  \fl(\bar B'_f(x))\cap \fl(\bar B'_f(y))$.   \eproof

This immediately implies, using that $\bar B_f(x)\subset \fl(\bar
B'_f(x))$ by Lemma~\ref{lem:beta}(ii), the following corresponding
floating version of Lemma \ref{lem:!}.

\begin{lemma}\label{FPlem:!} Let  $f\in S(\Hd)$,  $\eta \ge
\frac{\ab}{8\bD^2(n+1)\kappa(f)^2}$ and
$u\leq\frac{1}{\Oh\left(\bD^2n^{5/2}\kappa(f)^3(\log S+n^{3/2}
\bD^2\kappa(f)^2)\right)}$.
\begin{description}
\item[(i)]
 For each component $\fl(U)$  of $\fl(G'_\eta)$, there is a unique
zero $\zeta_U\in Z(f)$ such that  $\zeta_U\in Z(\fl(U))$. Moreover
$\zeta_U\in \cap_{x\in \fl(U)}\bar B_f(x)$.
\item[(ii)] If $\fl(U)$ and $\fl(V)$ are different components of $\fl(G'_\eta)$, then
$\zeta_U\ne \zeta_V$.\eproof
\end{description}
\end{lemma}

In order to show the correctness of {\tt Count\_Roots\_2}, we only
need to prove that $Z(f)\subset Z(\fl(G'_\eta))$. This  easily
follows adapting the proof of Part~(1) in Section~\ref{sec:proof1}
to this situation, making use of Lemma~\ref{FPlem:!} and the facts
that  Condition (i), $\fl(d(x_i,x_j))>\fl(\frac{3}{2}\pi
\eta\sqrt{n+1})$, implies that $d(x_i,x_j)>\pi\eta\sqrt{n+1}$
(Lemma~\ref{lem:e1}(ii)) and Condition (ii), $\fl(\|f(x)\|_\infty)>
\fl(\frac{\sqrt 2}{2}\pi \eta\sqrt{(n+1)\bD})$, implies that
$\|f(x)\|_\infty
>\frac{\pi}{2}\eta\sqrt{(n+1)\bD} $
(Lemma~\ref{lem:e2}(ii)).

\subsection{Proof of Theorem~\ref{th:main}(4): Complexity}\label{sec:comp}

We want to show that if $\eta\leq\frac{\ab}{4\bD^2(n+1)\kappa(f)^2}$
then {\tt Count\_Roots\_2}$(f)$ halts. Note that this means that
$$\frac{\ab}{8\bD^2(n+1)\kappa(f)^2}< \eta \le
\frac{\ab}{4\bD^2(n+1)\kappa(f)^2}$$   and hence, by
\S~\ref{sec:corr}, that it correctly returns $\#_{\R}(f)$.

Because of the hypothesis on $\eta$, the hypotheses of
Lemmas~\ref{lem:2pointsN}, and~\ref{lem:L2'} are satisfied.\\
Let $\fl(U)\ne\fl(V)$ be different components of $\fl(G'_\eta)$, and
therefore, by Lemma~\ref{FPlem:!}, $\zeta_U\ne \zeta_V$, and for all
$x\in \fl(U)$, $y\in \fl(V)$, by  Lemma~\ref{lem:2pointsN},
$d(x,y)>2\pi \eta\sqrt{n+1}$ holds.  This implies, by
Lemma~\ref{lem:e1}(i), that Condition~(i) in {\tt Count\_Roots\_2}
is satisfied.

Consider now $x\not\in\fl(A'(f))$. By
Proposition~\ref{prop:alphabar}(i),
$\bar\alpha(f,x)\geq\frac{\ab}{3}$. This implies, by
Lemma~\ref{lem:L2'}, that $\|f(x)\|_\infty >
\pi\eta\sqrt{(n+1)\bD}$, which in turn, by Lemma~\ref{lem:e2}(i),
ensures that Condition~(ii) in {\tt Count\_Roots\_2} is satisfied.
Hence, the algorithm halts.
\medskip

\noindent
{\bf Aknowledgement.} We are grateful to Andr\'e Galligo for a helpful
discussion.

{\small
%  \bibliography{../../../book/book}

}

\end{document}